\documentclass[11pt]{article}
\pdfoutput=1
\usepackage{jcapmod}
\usepackage{booktabs}
\usepackage[english]{babel}
\usepackage{amsmath,amssymb,amsbsy,amstext, amsthm, simplewick}
\usepackage{wrapfig}
\usepackage{graphicx}
\usepackage{amsfonts}
\usepackage{amssymb}
\usepackage{upgreek}
 \usepackage{exscale,relsize}
 \usepackage[makeroom]{cancel}
\usepackage{soul}
\usepackage{bbold}

\RequirePackage{color}

\usepackage{colortbl}
\definecolor{rp}{cmyk}{0.2, 1, 0.6, 0}
\definecolor{green2}{cmyk}{0, 1, 0.5, 0}
\definecolor{lightgreen}{cmyk}{0.2, 0, 0.2, 0.2}
\definecolor{turq}{cmyk}{1, 0, 0.2, 0.2}
\definecolor{lightgray}{cmyk}{0.1,0.2,0,0.1}
\definecolor{lightgray2}{cmyk}{0.4,0.4,0,0.8}
\definecolor{black}{cmyk}{1.0,1.0,1.0,1.0}

\allowdisplaybreaks[1]

\usepackage{colortbl}
\definecolor{lightgreen}{cmyk}{0.2, 0, 0.2, 0.2}
\definecolor{lightgray}{cmyk}{0.1,0.2,0,0.1}
\definecolor{lightgray2}{cmyk}{0.1,0.1,0,0.1}

\makeatletter
\newlength{\apb@width}
\newcommand{\autoparbox}[2][c]{\settowidth{\apb@width}{#2}\parbox[#1]{\apb@width}{#2}}

\makeatother

\numberwithin{equation}{section}

\def\d{{\rm d}}

\def\beq{\begin{equation}}
\def\eeq{\end{equation}}
\def\bea{\begin{eqnarray}}
\def\eea{\end{eqnarray}}

\def\Beq{\begin{equation}\begin{aligned}}
\def\Eeq{\end{aligned}\end{equation}}

\def\Ns{N_{\rm s}}
\def\Nf{N_{\rm f}}
\def\Nc{N_{\rm c}}

\def\M{{\sf M}}
\def\T{{\sf T}}
\def\t{{\sf t}}
\def\r{{\sf r}}
\def\n{{\sf n}}

\def\u{{\sf u}}
\def\v{{\sf v}}
\def\z{{\sf z}}
\def\L{{\sf \Lambda}}
\def\m{{\sf m}}
\def\Nf{{N_{\rm{f}}}}

\def\I{\mathbb{1}}
\DeclareRobustCommand{\SkipTocEntry}[4]{}

\begin{document}

\begin{titlepage}

\setcounter{page}{1} \baselineskip=15.5pt \thispagestyle{empty}

\bigskip\

\vspace{1cm}
\begin{center}

{\fontsize{20}{24}\selectfont  \sffamily \bfseries  From Wires to Cosmology}

\end{center}

\vspace{0.2cm}

\begin{center}
{\fontsize{13}{30}\selectfont  Mustafa A.~Amin$^{\clubsuit,\blacklozenge, \bigstar}$ and Daniel Baumann$^{\bigstar, \spadesuit}$}
\end{center}

\begin{center}

\vskip 8pt
\textsl{$^\clubsuit$ Physics \& Astronomy Department, Rice University, 6100 Main Street, Houston, U.S.A.}
\vskip 7pt

\textsl{$^ \blacklozenge$ Kavli Institute for Cosmology,
University of Cambridge, Madingly Road, Cambridge, U.K.}
\vskip 7pt

\textsl{$^\bigstar$ DAMTP, University of Cambridge, Wilberforce Road, Cambridge, U.K.}
\vskip 7pt

\textsl{$^\spadesuit$ Institute of Physics, University of Amsterdam, 1090 GL Amsterdam, The Netherlands}

\end{center}

\vspace{1.2cm}
\hrule \vspace{0.3cm}
\noindent {\sffamily \bfseries Abstract} \\[0.1cm]
We provide a statistical framework for characterizing stochastic particle production in the early universe via a precise correspondence to current conduction in wires with impurities.   Our approach is particularly useful when the microphysics is uncertain and the dynamics are complex, but only coarse-grained information is of interest. We study scenarios with multiple interacting fields and derive the evolution of the particle occupation numbers from a Fokker-Planck equation.  At late times, the typical occupation numbers grow exponentially which  is the analog of Anderson localization for disordered wires.
Some statistical features of the occupation numbers show hints of universality in the limit of a large number of interactions and/or a large number of fields.  For test cases, excellent agreement is found between our analytic results and numerical simulations.
\vskip 10pt
\hrule
\vskip 10pt

\vspace{0.6cm}
 \end{titlepage}

\tableofcontents

\newpage
\section{Introduction}
 
\begin{figure}[h!]
\centering
\includegraphics[scale=0.67]{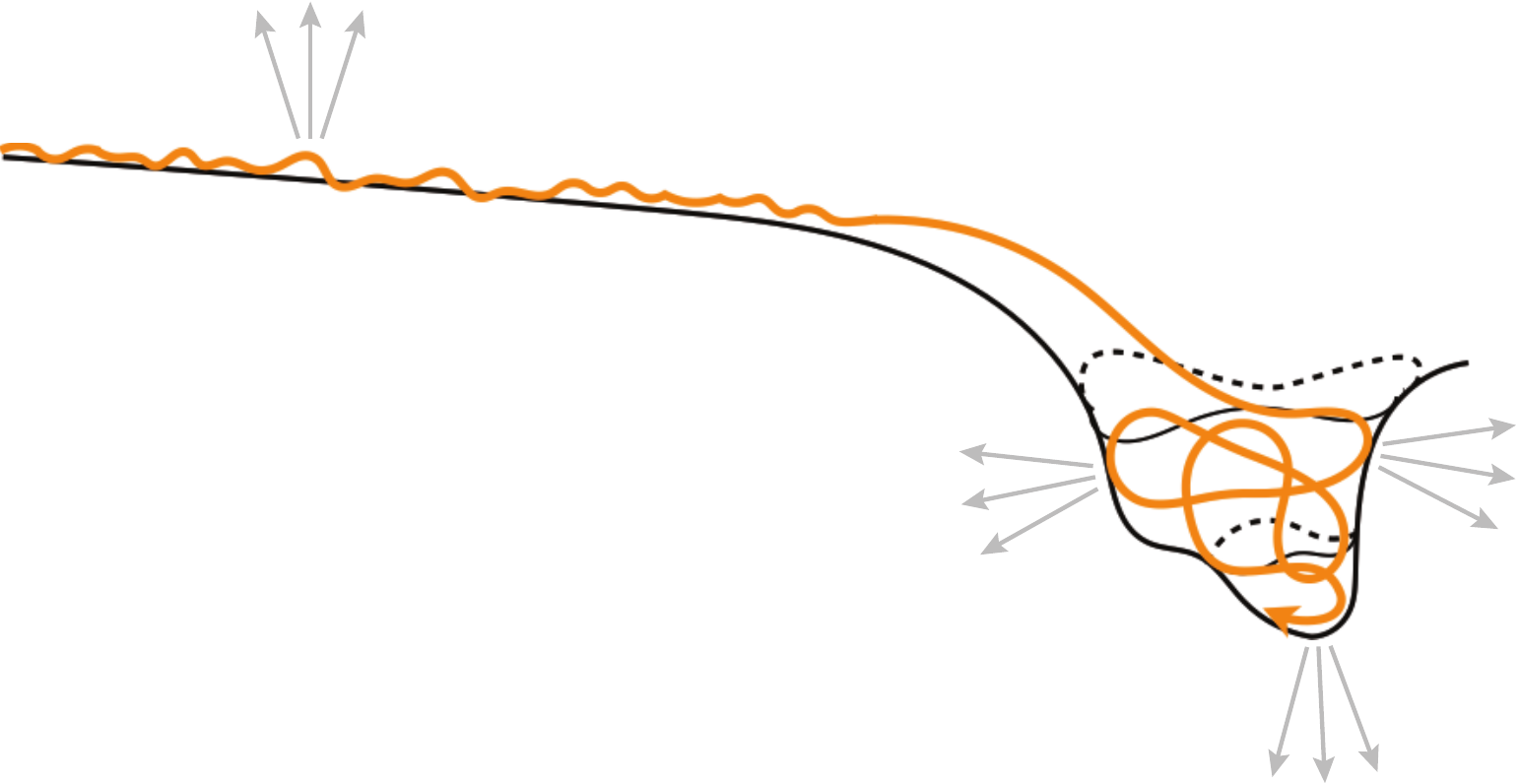}
\caption{The goal of this paper is to develop a framework for computing cosmological particle production in systems with a large number of random non-adiabatic events during and after inflation.}
\label{fig:dynamics}
\end{figure}

 In cosmology and particle physics we tend to be guided by the belief (or hope) that our theories become simpler at high energies: symmetries are restored and the number of degrees of freedom is reduced. 
 Although this reductionistic point of view \cite{1992dft..book.....W} has been fantastically successful in the development of the Standard Model, there is no guarantee that it also applies to the physics of the primordial universe.  In fact, recent attempts to find ultraviolet completions of models of inflation and reheating often are very complex, involving many fields and complicated interactions~\cite{Baumann:2014nda}. 
Analyzing such scenarios can be challenging~\cite{Tye:2008ef,Green:2009ds,Tye:2009ff,Braden:2010wd, Achucarro:2010jv, Achucarro:2010da, Chen:2011zf, Battefeld:2011yj, McAllister:2012am, Battefeld:2012wa, Greenwood:2012aj, Marsh:2013qca,Easther:2013rva,Hertzberg:2014iza,Watanabe:2015eia,Dias:2015rca,Figueroa:2015rqa,Jain:2015mma,DeCross:2015uza,Chluba:2015bqa,Amin:2014eta}, both due to insufficient information regarding the allowed theoretical constructions and due to limited constraints on model parameters from observations.  In some cases, the complexity of the microscopic description can lead to significant elements of randomness in the dynamics (e.g.~the masses and couplings of fields may fluctuate stochastically, reminiscent of disorder in condensed matter systems~\cite{Bassett:1997gb,Green:2014xqa}). Whenever the evolution is sufficiently non-adiabatic or tachyonic, it will involve significant amounts of stochastic particle production (see Fig.~\ref{fig:dynamics}). In this paper, we develop a framework to analyze such systems and study some simple toy models.  Applications to more realistic models of inflation and reheating will be presented in future work.  

Our study will be facilitated by a precise mathematical equivalence between stochastic particle production in cosmology and electron transport in wires.  A simplified version of the correspondence is as follows
\begin{equation}
\frac{d^2 \psi}{dx^2} +\left[E-V(x)\right]\psi = 0 \quad \longleftrightarrow \quad \frac{d^2 \chi_k}{d \tau^2} +\left[k^2+m^2(\tau)\right] \chi_k = 0\, , \label{equ:1}
\end{equation}
where, on the left-hand side, $\psi(x)$ is the electron wavefunction and $V(x)$ is a spatially-dependent potential due to the presence of random impurities. On the right-hand side, $\chi_k(\tau)$ is the mode function for a field in a time-dependent background and $m^2(\tau)$ is the effective mass with random non-adiabatic events arising from complicated interactions.\footnote{Equation (\ref{equ:1}) is easily generalized to multiple fields (multiple conduction channels) and more complicated couplings. Our framework only relies on there being a unitary map from some initial state to a final state, with random interactions in between.} 
The spatial dimension of the wire becomes time in the cosmological context, and the random impurities play the role of the stochastic time evolution of the effective mass. The correspondence  in (\ref{equ:1}) then allows us to apply many of the powerful tools developed by the condensed matter community~\cite{muller2010disorder, Beenakker:1997zz,janssen2001fluctuations, mello2004quantum} directly to the analogous cosmological problems.\footnote{The correspondence between random Schr\"odinger operators and particle production in the context of reheating was elegantly exploited in~\cite{Bassett:1997gb, Zanchin:1997gf,Zanchin:1998fj}.} It also means that known phenomena in the theory of disordered wires should have counterparts in the cosmological context.  One of the most dramatic effects of random disorder in wires is the exponential localization of the electron wavefunction, $|\psi(x)|^2 \sim e^{-|x|/\xi}$, where $\xi$ is the localization length.  
 Such {\it Anderson localization}~\cite{anderson1958absence} arises due to the interference of waves which are scattered by the impurities. By formulating cosmological particle production as a scattering problem, we will show\footnote{The connection between Anderson localization and particle production has been pointed out before, see for example~\cite{Hu:1997iu, Bassett:1999mt,Bassett:2005xm,Brandenberger:2008xc}. However, to the best of our knowledge, a detailed statistical formalism for understanding non-adiabatic particle production with multiple interacting fields and stochastic interactions has never been worked out. The present work was inspired by discussions of Anderson localization in \cite{Brandenberger:2008xc, Green:2014xqa}, albeit with different motivations and applications. In \cite{Tye:2007ja, Podolsky:2008du} an analogy was drawn between Anderson localization and inflation in higher-dimensional field spaces. This is unrelated to our proposal.} that Anderson localization maps to exponential particle production, $|\chi_k(\tau)|^2 \sim e^{+\mu_k \tau}$, where $\mu_k$ is the mean particle production rate.\footnote{Since space has been mapped to time, we are always in the one-dimensional situation where Anderson localization is particularly efficient.}  

\vskip 4pt
Crucially, the conduction properties of wires are determined by the statistics of the random impurities.  Analogously, stochastic particle production is characterized by the statistics of the non-adiabatic events. Taking conduction in disordered wires as an inspiration, we will develop a statistical framework for studying stochastic particle production in the early universe. 
Specifically, we will show that the occupation number of the produced particles, $n_k$, executes a drifting Brownian motion and derive a Fokker-Planck (FP) equation that evolves the probability distribution, $P(n_k;\tau)$. The precise structure of the FP equation is determined by the microscopic details of the scattering events.  Pleading maximum ignorance, we will use a {\it maximum entropy ansatz}~\cite{mello2004quantum} to parameterize this physics. We will show that the asymptotic solution to the FP equation is approximately a log-normal distribution and compute the evolution of the mean and the variance of the particle occupation number. 
The advantage of the statistical approach is that it reduces the complexity of the microscopic description to a few effective parameters of the coarse-grained theory (e.g.~$\xi$ and $\mu_k$).   In the context of inflation and reheating, focusing on coarse-grained characteristics is particularly relevant since both fundamental theory and cosmological observations are unlikely to provide  enough details about the relevant microphysics.

\vskip 4pt
Real wires of course aren't one-dimensional, but have finite cross sections. This allows a finite number of transverse modes of the wavefunction to be excited, which gives rise to coupled, longitudinal `conduction channels' (see e.g.~\cite{Beenakker:1997zz}).  We will show that multi-channel conduction maps to stochastic particle production with `multiple fields'.  
The output of the stochastic particle production is the joint probability distribution of the occupation numbers in each channel. We will derive a Fokker-Planck equation for this distribution function. (In the condensed matter context this is known as the DMPK equation~\cite{dorokhov1982transmission,mello1988macroscopic}.)  We will use this equation to compute the moments of the distribution and study them as a function of time and the number of fields.  In general, the rate of growth of the moments of the distribution depend on the number of fields in a way that we can predict.   In the limit of a large number of interactions and/or a large number of fields, we find interesting universality in the statistical distribution of the produced particles. For example,
the leading contribution in the variance of the total particle density is independent of the number of fields, a feature that we consider to be similar to the famous effect of `universal conductance fluctuations'\hskip 1pt\footnote{This refers to the fact that for weakly localized samples the fluctuations in the conductance for different samples are independent of the number of channels.}~\cite{altshuler1985fluctuations, lee1985universal} in multi-channel wires.  

\vskip 10pt
The outline of the paper is as follows. In Section~\ref{sec:Wires2Cosmo}, we develop the precise relationship between Anderson localization in disordered wires and stochastic particle production in cosmology. 
We give a derivation of the typical transmission probability of electrons in a wire and show that it maps inversely to the number of particles created in the cosmological context.
In Section~\ref{sec:Brownian}, we derive a Fokker-Planck equation describing the evolution of the probability density of the produced particles. We use this equation to determine the statistical properties of the particle production in detail.  In Section~\ref{sec:MultiField}, we generalize our treatment to scenarios with multiple fields. We show that this situation naturally maps to multi-channel conduction in wires and present a Fokker-Planck  description of such systems. In Section~\ref{sec:Conclusions}, we state our conclusions and outline our plans for future work.  A few technical details are relegated to the appendices:
In Appendix~\ref{sec:FP}, we derive the multi-field Fokker-Planck equation. In Appendix~\ref{sec:MEA}, we provide a maximum entropy analysis of the probability distribution. In Appendix~\ref{sec:QM}, we compute particle production in a few explicit examples where the transmission and reflection coefficients  for a single scattering can be obtained analytically~\cite{1965qume.book.....L}.

\vskip 6pt
Throughout, we will use 
natural units, $c=\hbar \equiv 1$.
The time variable will be $\tau$, and overdots will denote derivatives with respect to $\tau$.
We will use {\sf sans serif} font for matrices, e.g.~$\M, \T, \t, \r$.

\section{Stochastic Particle Production}
\label{sec:Wires2Cosmo}

Fields in a time-dependent background may have time-dependent couplings and effective masses.
Whenever the evolution is non-adiabatic, this leads to a burst of particle production~\cite{Kofman:1994rk,Shtanov:1994ce, Kofman:1997yn}.
To illustrate this, consider a scalar field $\chi$ with a time-dependent effective mass $m(\tau)$.
Let the linearized equation of motion of a single Fourier mode $\chi_{k}$ be
 \beq
\frac{d^2 \chi_{{k}}}{d \tau^2}  + \big[k^2 + m^2(\tau)\big]\hskip 2pt \chi_{{k}}=0\, .\label{KG}
\eeq
 In general, the equation of motion for $\chi_{{k}}$ may contain additional terms---e.g.~from couplings to other fields (see~Section~\ref{sec:MultiField})---but this will not lead to qualitative differences in our treatment. The mass term $m^2(\tau)$ may have an average adiabatic part (e.g.~due to the background FRW expansion), as well as a stochastic part with localized non-adiabatic events\footnote{In this context, non-adiabaticity means $|\dot{\omega}/{\omega^2}|\gg 1$, where $\omega^2(\tau) \equiv k^2+m^2(\tau)$.} (e.g.~due to the complex interactions in a higher-dimensional field space).  We wish to study the stochastic particle production in this situation. 
For simplicity, we will set the adiabatic piece of $m^2(\tau)$ to zero.   The evolution between the stochastic features will then be determined by plane wave solutions, $e^{\pm i k \tau}$.  An adiabatic contribution to $m^2(\tau)$ could be accounted for by replacing the plane wave mode functions by the exact solution (or its WKB approximation).

\begin{figure}[h!]
\centering
\includegraphics[scale=0.65]{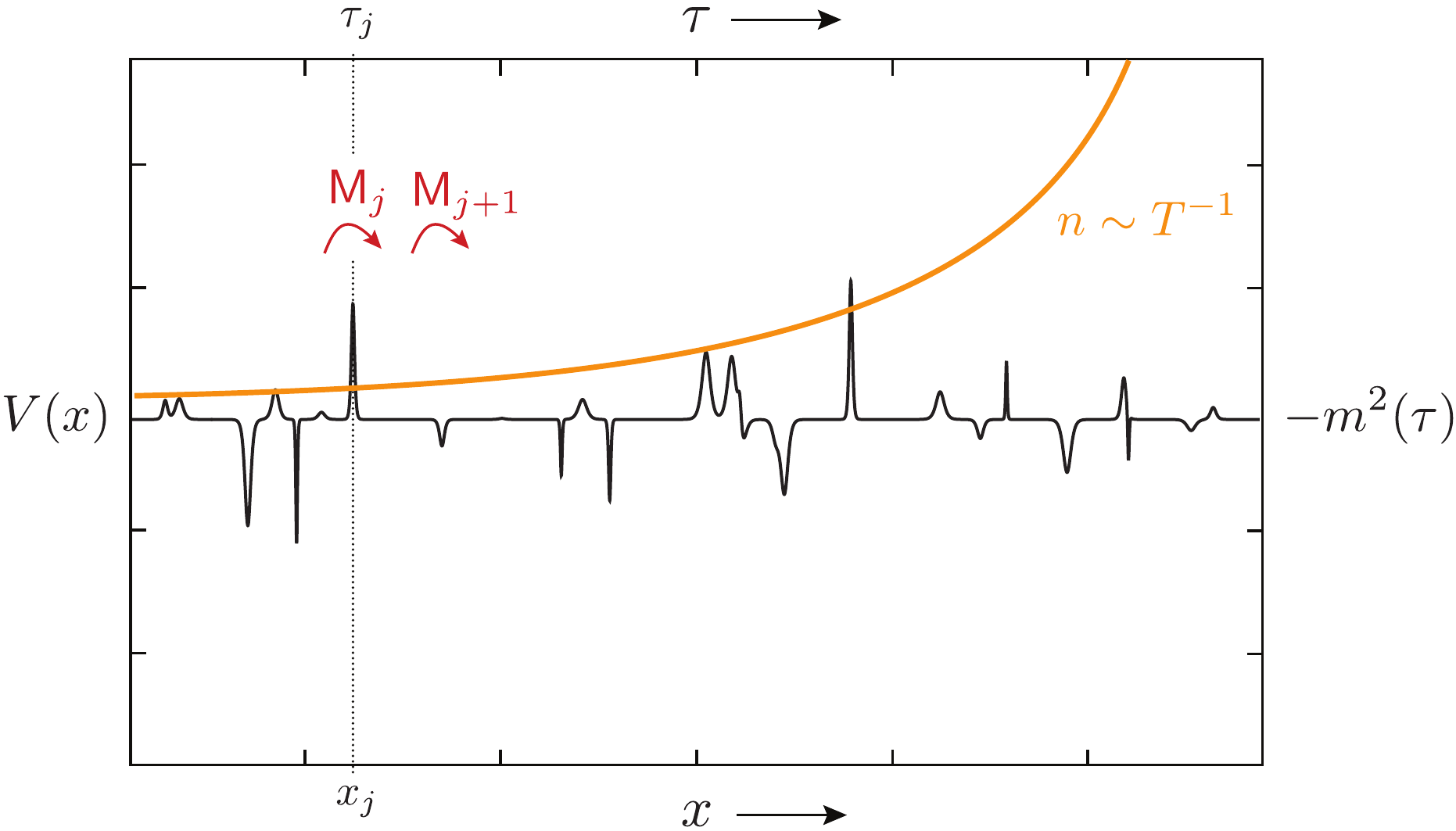}
\caption{Both the conduction of electrons in a wire and stochastic particle production in a cosmological scenario can be formulated as a scattering problem. The transfer matrices $\M_j$ describe the evolution across each scattering site (particle production event). The average total transmission probability $T$ decreases exponentially and the particle occupation number $n$ increases exponentially.}
\label{fig:correspondence}
\end{figure}

\subsection{Correspondence to Wires}
\label{sec:correspondence}

Consider a wire with random impurities. The flow of electrons in the wire will be strongly influenced by the presence of the impurities. For simplicity, we will ignore the spin of the electrons and consider them to be independent, i.e.~we don't take into account the interactions between the electrons.
 The time-independent Schr\"odinger equation for non-relativistic electrons of mass~$m_e$ then is
\beq
\left[- \frac{\hbar^2}{2m_e} \frac{d^2}{dx^2} + V(x) \right] \psi(x) = E\hskip 1pt \psi(x)\, ,\label{Schrodinger0}
\eeq 
where the potential $V(x)$ captures the effect of the random impurities (see Fig.~\ref{fig:correspondence}).
It will be convenient to set $2m_e/\hbar^2 \equiv1$ and write $E \equiv k^2$. The Schr\"odinger equation then becomes
\beq
\frac{d^2 \psi}{d x^2} + \big[k^2-V(x)\big]\hskip 2pt \psi=0\, . \label{S}
\eeq
It is easy to see that eqs.~(\ref{KG}) and (\ref{S}) map into each other if we exchange time and space, $\tau  \leftrightarrow x$, and make the identifications $\chi_{k}(\tau) \leftrightarrow \psi(x)$ and $m^2(\tau)  \leftrightarrow  - V(x)$.  
This shows that there is a precise mathematical correspondence between stochastic particle production and current conduction in wires.
We therefore expect to be able to translate many of the well-known results concerning conduction in wires to the cosmological context (see Table~\ref{tab:correspondence}).
 \begin{table*}[t]
 \begin{center}
 \begin{tabular}{ | c |l|  c |l|}
  \hline
  \multicolumn{2}{|c|}{\it Conduction in Wires} &  \multicolumn{2}{|c|}{\it Particle Creation in Cosmology}\\
  \hline
  \hline
  \ {\it Symbol}\ \ & \ {\it Meaning} &   \ {\it Symbol}\ \ & \ {\it Meaning}  \\
  \hline
  $x$ &\ distance & $\tau$ &\ (conformal) time \\
  $V(x)$ &\ potential & $- m^2(\tau)$ &\ negative mass-squared \\
  $\psi(x)$ &\ wave function & $\chi_{k}(\tau)$ &\ mode function \\[2pt]
  \hline
  $\Ns$ & \ number of scatterers &  $\Ns$ & \ number of non-adiabatic events\ \\
  $\Delta x$ & \ distance between scatterers\ \ & $\Delta \tau$ & \ time between non-adiabatic events\ \  \\
  $\rho$ & \ resistance & $n_k$ & \ particle occupation number  \\
  $\xi$ & \ localization length & $\mu_k$ & \ ``local" mean particle production rate\ \ \\
  \hline
  $\Nc$ & \ number of channels &   $\Nf$ & \ number of fields  \\
  \hline
  \end{tabular}
     \caption{Summary of the correspondence between physical quantities describing conduction in one-dimensional wires and those characterizing particle production in a time-dependent background.}
     \label{tab:correspondence}
     \end{center}
\end{table*}

\subsection{Conduction as a Scattering Problem}
\label{sec:scattering}
The conductance of the wire is related to the transmission probability of electrons across the wire~\cite{landauer1957spatial, imry1999conductance}, which can be obtained by solving the Schr\"odinger equation~(\ref{S}) in the presence of the impurities. 
\vskip 4pt
We begin by reviewing the scattering by a single impurity at $x=x_j$ (which will also help set up relevant notation and definitions).
To the left ($L$) and the right~($R$) of the impurity, 
the wavefunction can be written as a linear combination of right-propagating waves, $e^{ikx}$, and left-propagating waves, $e^{-ikx}$, 
\beq
\begin{aligned}
\psi_L(x)&=\beta_L \hskip 1pt e^{i k x}+\alpha_L  \hskip 1pt  e^{-i k x}\ , \\
\psi_R(x)&=\beta_R  \hskip 1pt  e^{i k x}+\alpha_R \hskip 1pt  e^{-i k x}\ . \label{equ:psi}
\end{aligned}
\eeq
The map between the state on the left and that on the right, can then be written as~\cite{1998qume.book.....M}
 \begin{equation}
\begin{pmatrix} \beta_R \\[2pt] \alpha_R \end{pmatrix}= 
\M_{j}
  \begin{pmatrix} \beta_{L} \\[2pt] \alpha_{L} \end{pmatrix} , \quad  \M_{j} \equiv\begin{pmatrix} 1/t^*_{j} & -r^*_{j}/t^*_{j} \\[2pt] -r_{j}/t_{j} & 1/t_{j} \end{pmatrix} , \label{equ:transfer}
\end{equation}
where ${\M}_{j}$ is called the {\it transfer matrix}, 
and $t_j$ and $r_j$ are complex transmission and reflection coefficients.  The transmission and reflection probabilities, $T_{j} \equiv |t_{j}|^2$ and $R_{j} \equiv |r_{j}|^2$, satisfy $R_j+T_j=1$. The form of the transfer matrix is fixed by unitarity and the reality of the potential.

\vskip 4pt
Ultimately, we want to chain several impurities together (see Fig.~\ref{fig:correspondence}). 
This is particularly easy to describe in the transfer matrix approach, since the total transfer matrix across $\Ns$ scatterers is simply given by the matrix multiplication of the individual transfer matrices:
\beq
{\M}(\Ns) \equiv {\M}_{\Ns} \ldots {\M}_{2} {\M}_{1}\, . \label{equ:MT}
\eeq
For notational convenience, we will often drop the argument $\Ns$ in the total transfer matrix $\M(\Ns)$. The total  transmission probability, and hence the conductance, can be obtained from $\M$.

\vskip 4pt
In the next section, we will give a short derivation \cite{muller2010disorder} of Anderson localization. In \textsection\ref{sec:PC}, we will show how this maps to exponential particle production.
\subsection{Anderson Localization}
\label{sec:Anderson}
Let us first consider the case of two adjacent impurities. 
The total transmission probability is then obtained from the 11-element of
the transfer matrix $\M= {\M}_{2} {\M}_{1}$. We find
\beq
T=\frac{T_{1} T_{2}}{\left|1-\sqrt{R_{1} R_{2}}\hskip 1pt e^{i \phi}\right|^2}\, , \label{trans2}
\eeq
where $\phi$ is the phase accumulated in the reflection between the two impurities. Note that this phase depends both on the separation between the impurities and their strengths. For simplicity, we will assume fixed strengths in this section, but our arguments do not change qualitatively if we relax this assumption. 
If the distance between the two impurities is random and uniformly distributed over a region with $k \Delta x\gg 1$, then the phase $\phi$ is also uniformly distributed between $0$ and $2\pi$. Taking the logarithm of the total transmission probability and averaging over the phase yields\hskip 1pt\footnote{We are imagining an ensemble of pairs of scatterers with varying separations, i.e.~an ensemble of different microscopic realizations of the disorder in the wire. In an experiment, we expect to measure a transmission probability which is appropriately averaged over many realizations of the disorder.}
\beq
\langle \ln T\rangle_\phi = \ln T_{1} +\ln T_{2} + \underbrace{\big\langle \ln |1-\sqrt{R_{1}  R_{2}}\hskip 1pt e^{i\phi}|^2\big\rangle_\phi}_{\displaystyle  \to\, 0}\ .
\eeq
We see that, after averaging over the phase, the logarithms of the transmission probabilities becomes additive (while the composition law for the transmission probabilities themselves is more complicated). 
The phase-averaged logarithm of the total transmission probability across $\Ns$ scatterers then simply is 
\beq
\langle \ln T(\Ns)\rangle_\phi  =\sum_{j=1}^{\Ns}\ln T_{j} \,\equiv\, - \Ns \gamma\, ,
\label{eq:logT}
\eeq
where $\gamma \equiv - \Ns^{-1}\sum_{j=1}^{\Ns}\ln T_{j}$ is sometimes referred to as the {\it Lyapunov exponent}.\footnote{This terminology highlights the analogy between the random scattering in wires and the stochastic time evolution in chaotic systems.}  We will also find it convenient to define the `typical' transmission probability as $T_{\rm typ} \equiv \exp[\langle \ln T\rangle_\phi]$.  This will correspond to the `most probable' transmission probability in the ensemble of random potentials. Using~(\ref{eq:logT}), we get
\beq
T_{\rm typ} = e^{-L/\xi}\,, \label{Tt} 
\eeq
where $L \equiv \Ns \Delta x$ is the total length of the wire and $\xi \equiv \Delta x/\gamma$ is the {\it localization length}. In one dimension, the localization length is of the same order as the transport mean free path~\cite{landauer1970electrical, thouless1973localization}.
  If the mean distance between scatterers, $\Delta x$, and the average logarithm of the transmission probability per scattering, $\gamma$, are fixed, then the total transmission probability decays exponentially with the length $L$ of the wire (or, equivalently, with the number of scatterers). This is Anderson localization~\cite{anderson1958absence}.  Naturally, the resistance of the wire scales inversely with the total transmission probability, $\rho_{\rm typ} \propto T_{\rm typ}^{-1}$, so the result~(\ref{Tt}) implies that $\rho_{\rm typ}$ grows exponentially with $L$.  At zero temperature, all one-dimensional wires are therefore insulators, independent of the strength of the impurities. 

\subsection{Particle Creation as a Scattering Problem}
\label{sec:PC}
The Klein-Gordon equation (\ref{KG}) can be solved in the same way as the 
Schr\"odinger problem, namely by formulating it as a scattering problem. In fact, the mapping to the treatment in \textsection\ref{sec:scattering} and \textsection\ref{sec:Anderson} is almost one-to-one.
The Fourier mode of the field after the $j$-th non-adiabatic event is
\beq
\chi_{j}(\tau)=\frac{1}{\sqrt{2k}}\left[\beta_{j} e^{ik\tau}+\alpha_{j} e^{-ik\tau}\right] , \label{equ:chi}
\eeq
where the overall normalization is chosen for future convenience.  To reduce clutter, we have suppressed the $k$-dependence of the mode functions and the Bogoliubov coefficients. The Wronskian, of the solutions, $W[\chi_j,\chi_j^*]$, is a constant. Consistency with vacuum initial conditions, $\beta_0=0$ and $\alpha_0=e^{i\delta}$, sets $W[\chi_j, \chi_j^*] = i$ and implies $|\alpha_j|^2-|\beta_j|^2=1$ for all $j$.

Now consider a single particle production event at $\tau =\tau_j$. In analogy with the scattering from an impurity, we relate the Bogoliubov coefficients before and after the non-adiabatic event by a transfer matrix 
 \begin{equation}
\begin{pmatrix} \beta_j \\[2pt] \alpha_j \end{pmatrix}\ =\  
\underbrace{\begin{pmatrix}  M_{11}& M_{12} \\[2pt] M_{21} & M_{22}\end{pmatrix}}_{\displaystyle \M_j}\,
  \begin{pmatrix} \beta_{j-1} \\[2pt] \alpha_{j-1} \end{pmatrix} .
  \label{equ:transfer1X}
\end{equation}
In practice, the elements of the transfer matrix $\M_{j}$ are determined by matching the solutions $\chi_j$ and $\chi_{j-1}$, and their derivatives, at $\tau=\tau_j$ (see Appendix ~\ref{sec:QM} for an example computation).  Using that $m^2(\tau)$ is real, the conjugate $\chi_j^*$ is also a solution to the equation of motion, which implies 
\begin{equation}
\begin{pmatrix} \alpha_j^* \\[2pt] \beta_j^* \end{pmatrix}\, =\, \M_j\,
  \begin{pmatrix} \alpha_{j-1}^* \\[2pt] \beta_{j-1}^* \end{pmatrix} .
    \label{equ:transfer2X}
\end{equation}
Comparing (\ref{equ:transfer1X}) and (\ref{equ:transfer2X}), we then find $M_{11} = M_{22}^*$ and $M_{12} = M_{21}^*$.
Moreover, since $|\alpha_j|^2 - |\beta_j|^2 = |\alpha_{j-1}|^2-|\beta_{j-1}|^2 = 1$, we have $|M_{11}|^2 - |M_{12}|^2 = 1$.
Defining $t_j\equiv 1/M_{11}^*$ and $r_j\equiv - M_{12}^*/M_{11}^*$, we can write the transfer matrix in the form of~\eqref{equ:transfer}:
\beq
\M_j =\begin{pmatrix}  M_{11}& M_{12} \\[2pt] M_{12}^* & M_{11}^*\end{pmatrix}=\begin{pmatrix}  1/t_j^*&-r_j^*/t_j^* \\[2pt] -r_j/t_j & 1/t_j\end{pmatrix} ,
\eeq
where $|r_j|^2 + |t_j|^2 = 1$.  Note that it wasn't necessary to define $r_j$ and $t_j$, but it makes the connection to the scattering problem particularly transparent.
 
 \vskip 4pt
As before, the total transfer matrix after $\Ns$ non-adiabatic events is the matrix multiplication of the individual transfer matrices ${\M}(\Ns) \equiv {\M}_{\Ns} \ldots {\M}_{2} {\M}_{1}$. After the $\Ns$-th scattering,  we have a negative frequency mode $\beta_{\Ns} e^{ik\tau}$, and a positive frequency mode $\alpha_{\Ns} e^{-ik\tau}$. The occupation number of a mode with frequency $k$ then is
\Beq
n(\Ns)
&\equiv\frac{1}{2k}\left(|\dot{{\chi}}_{\Ns^{\phantom \dagger}}|^2+k^2|{{\chi}_{\Ns^{\phantom \dagger}}}|^2\right)-\frac{1}{2}\,,\\[2pt]
&=\frac{1}{2}\left(|\alpha_{N_{\rm s}}|^2+|\beta_{N_{\rm s}}|^2\right)-\frac{1}{2}\,,\\[2pt]
&=|\beta_{N_{\rm s}}|^2\, .
\Eeq
For vacuum initial conditions, $(\alpha_0=e^{i\delta},\beta_0=0)$, this becomes
\Beq
n(\Ns)= |M_{12}(N_{\rm s})|^2 =\frac{|r(\Ns)|^2}{|t(\Ns)|^2}=T(\Ns)^{-1}-1\,.
\Eeq
We see that a large occupation number corresponds to a  small transmission probability in the equivalent scattering problem. 

In \textsection\ref{sec:Anderson}, we showed that $\ln T_{j}$ is additive after ensemble averaging and defined the typical transmission probability $T_{\rm typ}$ after many scatterings.  Repeating these arguments with $x\to \tau$ and  $T_j \to(1+n_j)^{-1}$, we arrive at
\Beq
\langle \ln(1+n)\rangle_\phi \,=\, \sum_{j=1}^{\Ns} \ln (1+n_j) \equiv \mu_k\tau\,,
\Eeq
where $\mu_k\equiv (\Ns\Delta\tau)^{-1}\sum_{j=1}^{\Ns} \ln (1+n_j)$ is the {\it mean particle production rate}.\footnote{Note that $n_j$ should be interpreted as {\it change} in the occupation number due to an {\it isolated} non-adiabatic event at $\tau=\tau_j$, whereas $n$ is the occupation number {\it after} $\Ns$ scatterings. Moreover, by assuming that the system is ergodic, we can interpret the ensemble average over microscopic realizations of the phase $\phi$ (determined  by the strengths and relative separations of the non-adaiabtic events) as being equivalent to an average over long times.}${}^{,}$\footnote{The dependence of $\mu_k$ on the wavenumber $k$ is determined by the details of the microphysics. For small $k$, one generically finds $\mu_k\propto k^{-2}$ (see e.g.~\cite{Bassett:1997gb} and Appendix~\ref{sec:QM}).} Note that the localization length and the particle production rate are inversely related $\xi \leftrightarrow \mu_k^{-1}$, and that both are determined by the dimensionless Lyapunov exponent: $\gamma_k =\mu_k \Delta\tau =  \Delta x/\xi$. The typical occupation number after many particle production events,
\Beq n_{\rm typ} \equiv \exp[\langle \ln(1+n)\rangle_\phi]-1\,,\Eeq
can  be related to the typical transmission probability of the equivalent scattering problem:
\beq
n_{\rm typ} = T_{\rm typ}^{-1}-1 = e^{+\mu_k \tau} -1\, . \label{equ:nt}
\eeq
The exponential behavior can also be understood as a Bose enhancement effect, i.e.~particle production is enhanced by existing particles in the mode.\footnote{We note that exponential growth doesnÕt arise if the fields are fermions \cite{Greene:1998nh,Peloso:2000hy}.} In the next section, we will arrive at this exponential behavior of the typical occupation number more formally. 
\begin{figure}[t!]
\centering
\includegraphics[scale=0.5]{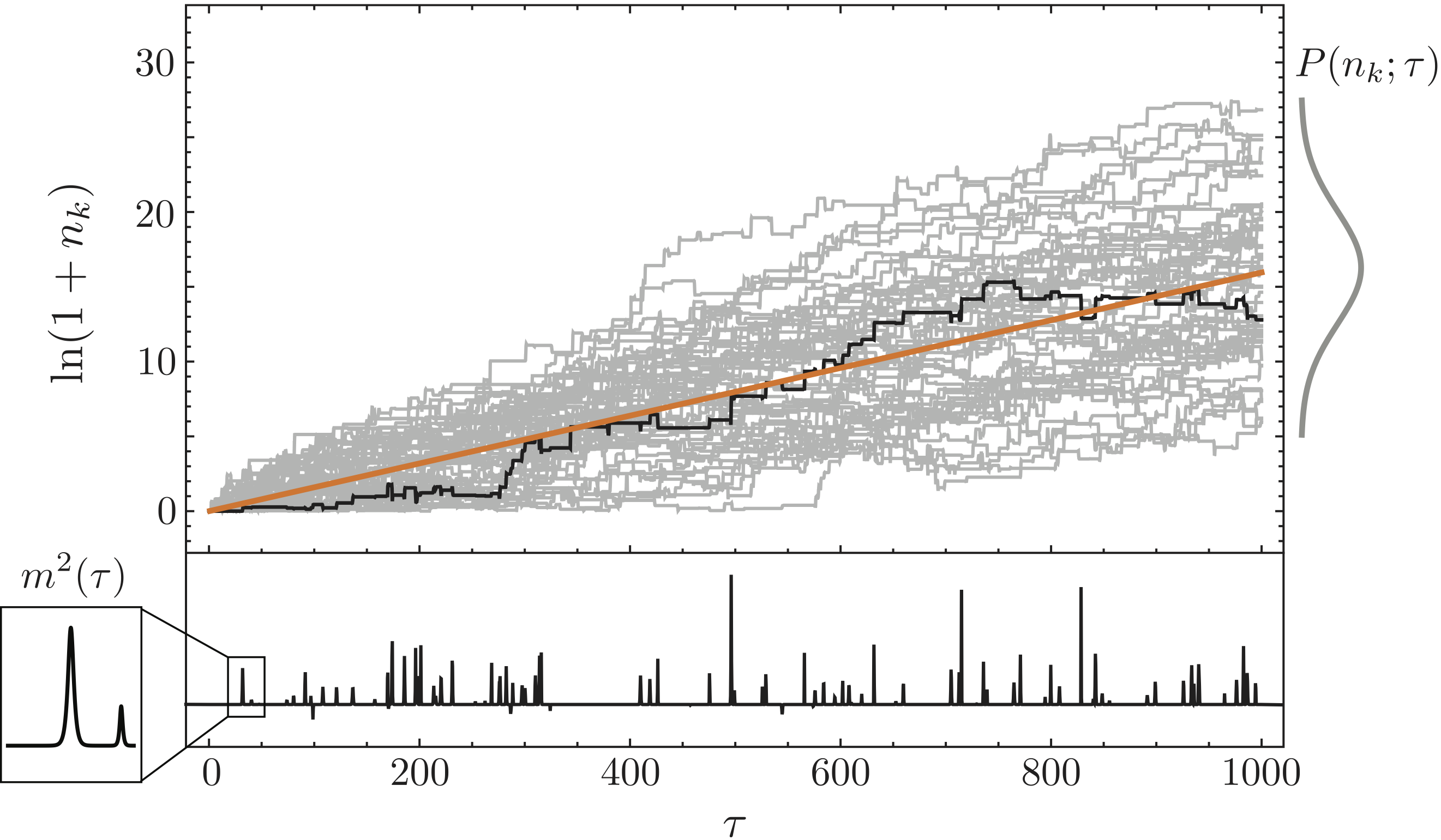}
\vspace{-0.2cm}
\caption{Evolution of the occupation number per mode in the presence of a large number of non-adiabatic interactions.  Each grey line shows the evolution for a given realization calculated numerically. The thick orange line is our analytic prediction for the most probable occupation number. The effective mass is modelled as $m^2(\tau)=\sum_{j=1}^{N_{\rm s}} \left({m_j}/{2w_j}\right)\, \textrm{sech}^2[(\tau-\tau_j)/w_j]$, where the widths $w_j$ and strengths $m_j$ are drawn from a Gaussian distribution. The locations $\tau_j$ are drawn from a uniform distribution in the interval of the simulation. The dimensionful variables $\tau$, $w_j$ and $m_j^{-1}$ are expressed in units of the inverse wavenumber $k^{-1}$. For the chosen values of the parameters, particle production mostly occurs due to violation of adiabaticity and only occasionally due to tachyonic instabilities (for details see Appendix~\ref{sec:QM}).} 
\label{fig:simulation}
\end{figure}

\vskip 4pt
We end this section by commenting on the striking difference in the solutions $\chi_k(\tau)$ and $\psi(x)$ of the same differential equation~\eqref{equ:1}, with $x\leftrightarrow \tau$ and $-V(x)\leftrightarrow m^2(\tau)$. As suggested by the growth in the occupation number,  the mode function $\chi_k(\tau)$ grows exponentially with time.
On the other hand, the wave function $\psi(x)$ decays exponentially with distance as suggested by the transmission probability for electrons in a wire. This apparent discrepancy can be understood as a difference between the initial condition for the Klein-Gordon equation and the boundary condition of the Schr\"odinger equation.

First, note that while it is natural in the cosmological context to evolve forward in time, 
in the case of the wire, 
we need to pick a spatial direction to define growth or decay. 
In practice, the symmetry $x \leftrightarrow -x$ is broken by an applied voltage; we will pick $x>0$ as the direction of the voltage drop.  
Second, in the time-dependent case it is natural to chose the vacuum solution $\chi_k\sim e^{-ik\tau}$ as the {\it initial} condition at $\tau=0$. Similarly, in the time-independent case it is natural to impose an outgoing {\it boundary} condition, $\psi\sim e^{+ikx}$, at $x=L$. For $\tau >0$, the mode function $\chi_k$ grows as it encounters scatterers, i.e. there is particle production. Similarly, for $x < L$, the wavefunction $\psi$ also ``grows" as one moves toward $x=0$.  In the direction defined by the voltage drop this corresponds to a decay of $\psi$.\
\section{Brownian Motion}
\label{sec:Brownian}

Equation~(\ref{equ:nt}) provides rather rudimentary information about the typical rate of particle production. We would like to obtain a more detailed understanding of the statistics of the produced particles as a function of time.
Figure~\ref{fig:simulation} shows a numerical solution for the evolution of the occupation number $n_k(\tau)$ for an ensemble of randomly generated mass functions $m^2(\tau)$.  
 We see that the function $n_k(\tau)$ executes a {\it drifting random walk}.  The occupation number after a time $\tau$ will be a stochastic quantity.  By considering how the system responds to `adding' a differential time interval $\delta \tau$ (and averaging over the randomness it contains), we can derive a Fokker-Planck (FP) equation for the evolution of the probability density, $P(n_k;\tau)$.\footnote{FP equations have also been used to describe the evolution of fluctuations in stochastic inflation~\cite{Salopek:1990re, Starobinsky:1986fx,Burgess:2014eoa}. In that case, the stochasticity in the equations of motion for long-wavelength fluctuations arises due to short-wavelength quantum fluctuations.} Armed with the FP equation, we can study the statistics of the produced particles as a function of time. 

\subsection{Fokker-Planck Equation}
\label{subsec:FPEq}
In this section, we derive the FP equation for stochastic particle production of a single field. A similar derivation of an FP equation for transmission probabilities in disordered wires was presented in~\cite{mello2004quantum}. The derivation is a bit lengthy, so the impatient reader may jump directly to the final answer~(\ref{equ:FP}) without loss of continuity.  
 \begin{figure}[h!]
\centering
\hspace{0.5cm}\includegraphics[scale=0.43]{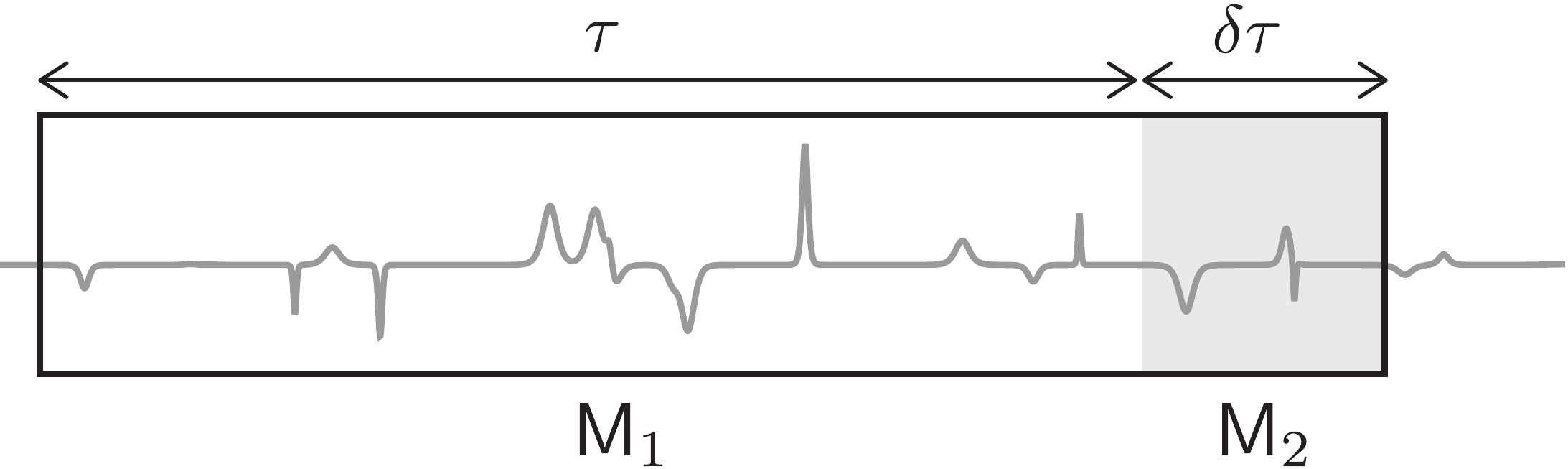}
\caption{A time interval of length $\tau$ and total transfer matrix $\M_1$ is followed by an infinitesimal interval~$\delta \tau$ with transfer matrix $\M_2$.  The transfer matrix of the combined system is $\M = \M_2 \M_1$.}
\label{fig:FP}
\end{figure}
 \vskip 4pt
We start by adding a small time interval $\delta \tau$ to an existing interval $\tau$ (see Fig.~\ref{fig:FP}). The transfer matrix for the elongated interval, $\M \equiv \M_{\tau+\delta \tau}$, can be written in terms of $\M_1 \equiv \M_\tau$ and $\M_2 \equiv \M_{\delta \tau}$, via the composition law $\M = \M_2 \M_1$.  The probability density of the total transfer matrix $\M$ can  formally be written as  the {\it Smoluchowski equation}\hskip 1pt\footnote{Equation (\ref{equ:SmolM}) only relies on the assumption that the process is Markovian.  In words, the probability of being here at a given time is equal to the probability of being somewhere else a bit earlier multiplied by the probability of making the transition from somewhere else to here (integrated over all places from which one can transition to here). Note that this part is not restricted to the single-field case (see Appendix~\ref{sec:FP}).}
\beq
P(\M;\tau+\delta \tau)=\int P(\M_1;\tau)\hskip 1pt P(\M_2;\delta\tau)\,\d\M_{2}\, \equiv\,\langle P(\M_1;\tau)\rangle_{\M_2}\, ,
\label{equ:SmolM}
\eeq
where $\M_1 = \M_2^{-1} \M$.  Writing $\M_1 \equiv \M+\delta \M(\M,\M_2)$, we can Taylor expand both sides of eq.~\eqref{equ:SmolM}:
\Beq
\partial_\tau P(\M;\tau)= \frac{\langle \delta \M\rangle_{\M_2}}{\delta \tau} \hskip 1pt \partial_\M P(\M;\tau)+ \frac{\langle \delta \M \delta \M\rangle_{\M_2}}{\delta \tau} \hskip 1pt\partial_\M \partial_\M P(\M;\tau)+\cdots\, .
\Eeq
This will become the FP equation for the occupation number $n$ after an appropriate parametrization of the transfer matrices and a marginalization over certain parameters. 

\vskip 4pt
 It will be convenient to write the transfer matrix \eqref{equ:transfer} in polar form~\cite{1988AnPhy.181..290M} (cf.~\textsection\ref{subsec:Matrix}):
\Beq
\M& =\begin{pmatrix} e^{i\theta}\sqrt{ 1+n}& \ e^{i(2\phi-\theta)}\sqrt{n} \\[2pt] e^{-i(2\phi-\theta)}\sqrt{n} & \ e^{-i\theta}\sqrt{1+n} \end{pmatrix}, \label{equ:polar}
\Eeq
where we defined
\Beq
 t& = \sqrt{T}e^{i\theta}\,,\\
r&=- \sqrt{1-T}e^{2i(\theta-\phi)}\,,\\ 
n&=T^{-1}-1\,. \label{equ:defs}
\Eeq
All quantities in (\ref{equ:polar}) and (\ref{equ:defs}) depend on the wavenumber $k$. We suppress this dependence in order to reduce clutter. 
The Smoluchowski equation \eqref{equ:SmolM} then becomes
\Beq
P(\{n,\theta, \phi\};\tau+\delta \tau)&=\int P(\{n_1,\theta_1, \phi_1\};\tau)P(\{n_2,\theta_2,\phi_2\};\delta \tau)\, \d n_2\frac{\d\phi_2}{2\pi}\frac{\d\theta_2}{2\pi}\\[4pt]
&\equiv\langle P(\{n_1,\theta_1,\phi_1\};\tau)\rangle_{\delta \tau}\, .
\label{equ:SmolP}
\Eeq
To be able to Taylor expand the right-hand side, we first write
$\{n_1,\theta_1, \phi_1\}$ in terms of $\{n,\theta,\phi\}$ and $\{n_2,\theta_2,\phi_2\}$.  This follows directly from the polar form of the transfer matrices and the relation~$\M_1 = \M_2^{-1}\M$. 
For example, we get 
\begin{align}
n_1 &=\left[{\M_1}\right]_{11}^*\left[{\M_1}\right]_{11}-1 \quad \ \ \, \equiv\, n+\delta n\, , \label{equ:n1sf} \\[2pt]
\label{equ:theta1sf}
\theta_1&= - \frac{i}{2} \ln([\M_1]_{11}/[\M_1]_{11}^*) \equiv \theta+\delta\theta \, ,
\end{align}
where 
\Beq
\label{equ:delta n}
\delta n \equiv n_2(1+2 n)-2\sqrt{(1+n_2)(1+n)n_2 n}\cos \left[2(\phi_2-\phi)\right] .
\Eeq
We note that $\delta n$ only depends on $\tilde{\phi}_2\equiv \phi_2-\phi$. The explicit expression for $\delta \theta$ will not be important; we will only need to know that it also only depends on $\tilde{\phi}_2$ and is independent of $\theta$.

\vskip 4pt
To make further progress, we need to make some physical assumptions about the form of the probability distribution of the transfer matrix in the small interval $\delta\tau$: $P_2  \equiv P(\{n_2,\theta_2,\phi_2\};\delta \tau)$.
We will be conservative and determine $P_2$ by the condition that is maximizes the Shannon entropy, $\mathcal{S}= - \langle \ln P_2\rangle_{\delta\tau}$, subject to certain constraints. The constraints  may be based on symmetry arguments, consistency requirements and available information regarding the microphysics. Following \cite{mello2004quantum}, we will refer to this as the {\it maximum entropy ansatz}. The minimal set of constraints we chose to include are: 
\begin{itemize}
\item[{\it i}\hskip1pt)] We assume that the local mean particle production rate is known. This means that we will~fix  
\Beq
\frac{\langle n_2\rangle_{\delta\tau}}{\delta\tau} \equiv \mu\,,
\Eeq
assuming that this quantity is calculable from the microphysics. 
\item[{\it ii}\hskip 1pt)] We require that $\M_{\tau+\delta\tau}\rightarrow\M_{\tau}$ in the limit $\delta \tau \to 0$. This seems eminently reasonable. It just means that the addition of an infinitesimal interval cannot lead to a finite change in the transfer matrix. 
\end{itemize}
In Appendix~\ref{sec:MEA}, we show that
these two constraints imply that $P_2 \to P(\{n_2,\theta_2\};\delta \tau)$, i.e.~there is no dependence on $\phi_2$.  For weak scattering, this corresponds to the scattering events being uniformly distributed. In what follows, the particular functional form of $P_2(\{n_2,\theta_2\};\delta\tau)$ imposed by the maximum entropy ansatz will not be important.\footnote{Our numerical simulations are not restricted to $P_2$'s which are consistent with the maximum entropy ansatz. The results, however, are consistent with the solutions of the FP equation. The fact that, in the limit of a large number of scatterings, the results become insensitive to $P_2$ is a consequence of the Central Limit Theorem (cf.~\textsection \ref{sec:Anderson} and~\textsection \ref{sec:PC}.).} The derivation of the FP equation is more general.

\vskip 4pt
Assuming simply that $P_2$ is independent of $\phi_2$, the Smoluchowski equation undergoes a dramatic simplification. 
First, we note that the distribution function satisfies the {\it persistence property}: if $P_{1,2}$ are independent of $\phi_{1, 2}$, then so is $P$, i.e.~$P(\{n,\theta, \phi\};\tau+\delta\tau)=P(\{n,\theta\};\tau+\delta\tau)$.  The proof is by induction.\footnote{Consider eq.~(\ref{equ:SmolP}) and recall that $n_1$ and $\theta_1$ are only functions of $\tilde \phi_2 = \phi_2-\phi$. By a change of variables the integral over $\phi_2$ then becomes an integral over $\tilde \phi_2$ and the dependence on $\phi$ disappears.} The Smoluchowski equation (\ref{equ:SmolP}) then becomes
\Beq
P(\{n,\theta\};\tau+\delta \tau)&= \langle P(\{n+\delta n,\theta + \delta \theta\};\tau)\rangle_{\delta \tau}\,.
\label{equ:SmolPwoPhi}
\Eeq
Integrating both sides with respect to $\theta$, we get
\Beq
P(n;\tau+\delta \tau) = \langle P(n+\delta n;\tau)\rangle_{\delta \tau}\,,
\Eeq
where we have used that $\delta\theta$ is independent of $\theta$. We apologize for the somewhat ambiguous notation: the $P$'s without the arguments $\theta$ should be understood as the original $P$'s integrated over $\theta$. Taylor expanding the left-hand side with respect to $\delta\tau$ and the right-hand side with respect to~$\delta n$, we find
\beq
\begin{aligned}
\frac{\partial}{\partial \tau} P(n;\tau)=\frac{\partial}{\partial n}P(n;\tau)\frac{\langle\delta n\rangle_{\delta \tau}}{\delta \tau}+\frac{1}{2}\frac{\partial^2}{\partial n^2}P(n;\tau)\frac{\langle(\delta n)^2\rangle_{\delta \tau}}{\delta \tau}+\cdots\, ,
\end{aligned}
\eeq
where, using \eqref{equ:delta n}, we have
\begin{align}
\langle\delta n\rangle_{\delta\tau}&=(\mu \hskip 1pt\delta\tau) (1+2 n)\,,\\
\langle(\delta n)^2\rangle_{\delta\tau}&=(\mu \hskip 1pt \delta\tau) \hskip 1pt 2n(1+ n)+\mathcal{O}[(\mu\delta\tau)^2]\,. \label{eq:dn}
\end{align}
In order to truncate the expansion in (\ref{eq:dn}) at lowest order in $\mu\hskip 1pt \delta \tau=\langle n_2\rangle_{\delta \tau}$, we require that  the local particle production rate is small.
This is the most limiting assumption of this derivation and should be kept in mind while applying our framework.

\begin{figure}[t!]
\centering
\includegraphics[scale=0.6]{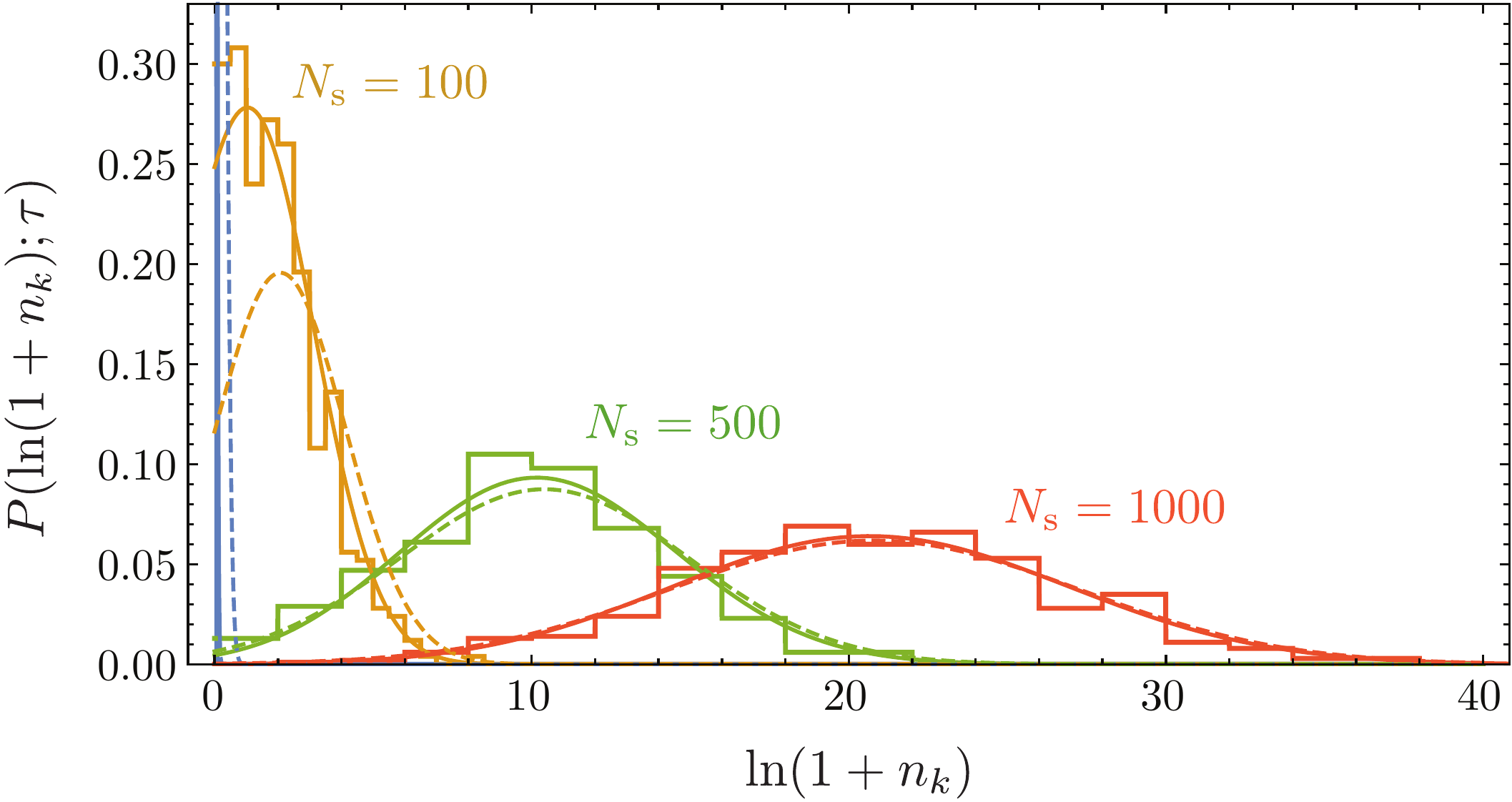}
\caption{Evolution of the probability density of the logarithm of the occupation number per mode, $\ln (1+n_k)$, as a function of time (or the number of scatterings $\Ns$). The agreement between our numerical simulations (solid lines of the histograms) and the exact solution (solid lines) is always excellent. The gaussian approximation (dashed lines) improves as the number of scatterings $\Ns$ becomes large.  }
\label{fig:Histogram}
\end{figure}

\vskip 4pt
Putting everything together, we arrive at  the final form of the Fokker-Planck equation
 \beq
\boxed{\ \frac{1}{\mu_k} \frac{\partial }{\partial \tau} P(n;\tau) = \underbrace{(1+2n)\frac{\partial}{\partial n}P(n;\tau)}_{\rm drift} \ + \ \underbrace{n(1+n) \frac{\partial^2}{\partial n^2} P(n;\tau)}_{\rm diffusion}\   } \ , \label{equ:FP}
 \eeq
 where we have restored the momentum dependence in the mean particle production rate, $\mu=\mu_k$, but left it implicit in the occupation number, $n_k = n$.   The FP equation (\ref{equ:FP2}) has an exact solution~\cite{muller2010disorder} for all $n$, although the integral form of the solution isn't very instructive.   In Fig.~\ref{fig:Histogram} we show the evolution of the probability as a function of time.  We find very good agreement between the result of our numerical simulations and the solution of the FP equation.\footnote{We used the transfer matrix approach of \textsection {\ref{sec:PC}} to solve ~\eqref{KG} numerically with $m^2(\tau)=\sum_{j=1}^{\Ns} m_j\delta_D(\tau-\tau_j)$.  For a single scattering, the transfer matrix is known analytically (see Appendix~\ref{sec:QM}). After drawing the locations~$\tau_j$ and amplitudes $m_j$ from a distribution, we calculated the occupation number by chaining together $\Ns$ transfer matrices.}

\vskip 4pt
It is also instructive to consider the asymptotic limit of the FP equation. For this purpose, let us first  write (\ref{equ:FP}) in the following form
 \beq
 \frac{1}{\mu_k} \frac{\partial }{\partial \tau} P(n;\tau) = \frac{\partial}{\partial n} \left(n (n+1) \frac{\partial P}{\partial n}\right) . \label{equ:FP2}
 \eeq
For $n \gg 1$, we have $n(n+1) \to n^2$ on the right-hand side and it is easy to show that the solution converges to the log-normal distribution
 \beq
 P(n;\tau)\hskip 1pt \d n = \frac{1}{\sqrt{4\pi \mu_k \tau}} \exp\left[-\frac{\big(\ln n-\mu_k \tau\big)^2}{4\mu_k \tau}\right] \d 
 \ln n\, . \label{equ:logP}
 \eeq
 This shouldn't be surprising. In \textsection\ref{sec:PC}, we saw that the phase-average of $\ln n$ is the sum of the logarithms of the particle occupation numbers produced at each scattering. The central limit theorem then suggests that $\ln n$ is Gaussian distributed (and $n$ obeys a log-normal distribution). This is true, except for deviations at small $n$.  These deviations arise because the total transmission probability is bounded by 1 (and $n$ is bounded by $0$).   

\subsection{Moments of the Density}
\label{ssec:moments}

 Although the solution of FP equation contains all the information about the statistics of the process, it is also convenient to instead look at the evolution of the moments of the occupation number directly.
The equation for the evolution of $\langle F(n)\rangle \equiv \int \d n \, F(n) P(n;\tau)$ can be easily obtained from the FP equation, and is given by
\begin{align}
 \frac{1}{\mu_k} \frac{\partial \langle F\rangle}{\partial \tau} &\,=\, \left\langle (1+ 2 n) \frac{\partial F}{\partial n} +  n(1+n) \frac{\partial^2 F}{\partial n^2}  \right\rangle \, .  \label{equ:F1}
\end{align}
By picking the functional $F$ conveniently, we can study arbitrary moments of the occupation number~$n$. In general, this leads to a set of coupled differential equations.   From (\ref{equ:F1}) it is easy to see that $\langle n \rangle$ and $\langle n^2 \rangle$ satisfy
 \begin{align}
\frac{1}{\mu_k}  \frac{\partial \langle n \rangle}{\partial \tau} &\,=\, 1 + 2\langle n \rangle\, , \label{equ:n1S} \\
\frac{1}{\mu_k}\frac{\partial \langle n^2 \rangle}{\partial \tau} &\,=\, 4 \langle n \rangle + 6 \langle n^2 \rangle\, . \label{equ:n2S}
\end{align}
Defining $\tau=0$ to be the time at which both $\langle n \rangle$ and  $\langle n^2 \rangle$ are vanishingly small, we get
\begin{align}
\langle n \rangle &= \frac{1}{2}\left(e^{2 \mu_k \tau} - 1 \right) , \label{equ:mean} \\ {\rm Var}[n] \equiv \langle n^2 \rangle- \langle n \rangle^2 &= \frac{1}{12}\left(1- 3 e^{4 \mu_k \tau} +2 e^{6\mu_k \tau} \right)  \ \xrightarrow{\ \mu_k \tau \gg 1\ } \ \frac{2}{3} \hskip 1pt e^{2\mu_k \tau} \langle n \rangle^2\, . \label{equ:var}
\end{align}
We see that the variance of the occupation number grows faster than the square of the mean. This illustrates that the probability density $P(n;\tau)$ becomes a very broad function at late times (cf.~Fig.~\ref{fig:Histogram}). The mean $\langle n \rangle$ is therefore not a good measure of the typical number of particles produced.

\vskip 4pt
To derive the evolution of the typical density, $ n_{\rm typ} \equiv  \exp[\langle \ln(1+n)\rangle]-1\,$, we consider the expectation value of $\ln(1+n)$ (and its higher-order moments): 
\begin{align}
\frac{1}{\mu_k}\frac{\partial \langle \ln(1+n) \rangle}{\partial \tau} &\,=\, 1 \, , \label{equ:logn1S} \\[2pt] 
\frac{1}{\mu_k}\frac{\partial \langle [\ln(1+n)]^2 \rangle}{\partial \tau} &\,=\, 2 \langle \ln(1+n) \rangle + 2 \langle n(1+n)^{-1} \rangle \, . \label{equ:logn2S}
\end{align}
The first equation can be integrated directly to give
 \beq
 \langle \ln(1+n)\rangle =  \mu_k \tau \quad \rightarrow \quad n_{\rm typ}  \,=\, e^{\mu_k \tau}-1 \, .  
\label{equ:nt2} 
\eeq
This is consistent with the result of our more heuristic derivation; cf.~eq.~\eqref{equ:nt}.
In the limit of late times, $\mu_k\tau \gg 1$, the last term in (\ref{equ:logn2S}) becomes $2\langle n(1+n)^{-1} \rangle \to 2$ and the system of equations closes.  Substituting (\ref{equ:nt2}) and integrating, we find
 \Beq
 \Delta {\rm Var}[\ln(1+n)]  = 2 \mu_k (\tau-\tau_0)\, , \label{equ:VarS}
 \Eeq 
where $\Delta f \equiv f-f_0$ and the subscript `0' denotes a quantity evaluated at the time $\tau_0$ (with $\mu_k\tau_0\gtrsim 1)$.
We see that the variance of $\ln(1+n)$ grows slower than the square of the mean:
\beq
\frac{\Delta {\rm Var}[\ln(1+n)]}{(\Delta \langle \ln(1+n)\rangle)^2} = \frac{2}{\mu_k (\tau-\tau_0)}\, .
\eeq
The mean of $\ln(1+n)$ is therefore a good measure of the number of particles produced (see Fig.~\ref{fig:simulation}). 
These results are consistent with the properties of the log-normal distribution~(\ref{equ:logP}).

\newpage
 \section{Generalization to Multiple Fields}
 \label{sec:MultiField}

Ultimately, one of our motivations is to describe the complex multi-field dynamics that may have occurred in the early universe. 
This also has a direct analog in the theory of disordered wires.  So far, we have ignored the finite thickness of the wire.  Taking the thickness  into account leads to a finite number of transverse excitations in the electron wavefunction.  This then gives rise to coupled, longitudinal `conduction channels'.  In this section, we will develop the framework of stochastic particle production with multiple fields and its correspondence to multi-channel conduction.

\subsection{Preliminaries}
\label{sec:Prelim}
Consider the action of $\Nf$ coupled scalar fields $\phi^a$,
 \beq
 S = \int \d^4 x \sqrt{-g}\left[\frac{M_{\rm pl}^2}{2}R - \frac{1}{2}G_{ab}(\phi^c)  \partial^\mu \phi^a \partial_\mu \phi^b - V(\phi^c) +\cdots\right] ,
 \eeq
 where $a,b,c = 1, \ldots, \Nf$.  The linearized equation of motion for the field fluctuations can be written in the following from (see e.g.~\cite{Amin:2014eta})
 \beq
\underbrace{ \left[ \mathbb{1}\, (\partial_\tau^2+k^2) + {\sf p}(k,\tau)\partial_\tau + {\sf m}(k,\tau) \right]}_{\displaystyle \equiv {\sf w}(k,\tau)} \cdot \,\delta \vec{\phi}_{k} \,=\, 0\ , \label{equ:EOM}
 \eeq
 where $\delta \vec{\phi}$ is a vector made out of the fluctuations of the fields $\phi^a$. The coefficient functions $ {\sf p}(k,\tau)$ and ${\sf m}(k,\tau)$ are matrices, with
 \begin{align}
 ({\sf p})^a{}_b &= 2 {\cal H} \delta^a_b + \cdots \, , \quad
  ({\sf m})^a{}_b = a^2 G^{ac} V_{,cb} + \cdots\, . \label{equ:PF}
 \end{align}
 The ellipses in (\ref{equ:PF}) stand for a complicated set of terms arising, for instance, from a nontrivial field space metric $G_{ab} \ne \delta_{ab}$. The precise form of (\ref{equ:EOM}) will not be important. All we care about here is that it defines a linear map describing the unitary evolution of $\delta \vec{\phi}_{k}(\tau)$. 
 For simplicity and concreteness, we will assume that\footnote{The assumption $G_{ab} =\delta_{ab}$ can also be justified from an effective field theory perspective~\cite{Green:2014xqa}: in cases of strong disorder in the mass term, the corrections to $G_{ab}$ are often irrelevant in the technical sense.} $G_{ab} = \delta_{ab}$, and ignore the Hubble expansion for the remainder of this section, i.e.~we set $({\sf p})^a{}_b =0$ and $({\sf m})^a{}_b =\delta^{ac}V_{,cb}$. We will refer to ${\sf m}$ as the {\it mass matrix}.
 
\vskip 4pt
We assume that the evolution of the field fluctuations contains localized non-adiabatic events at random intervals around $\tau = \tau_j$, and that the fields are otherwise free.
After the $j$-th event, the evolution of the fields is given by 
\Beq
\delta \vec{\phi}_j(\tau)=\frac{1}{\sqrt{2k}}\left[\vec{\beta}_je^{ik\tau}+\vec{\alpha}_je^{-ik\tau}\right] ,
\Eeq
where we have suppressed the dependence of $\delta\vec{\phi}_j$ and $(\vec{\alpha}_j,\vec{\beta}_j)$ on $k$ to reduce clutter. The Bogoliubov coefficients before and after the non-adiabatic event are related by 
 \begin{equation}
\begin{pmatrix} \vec{\beta}_j \\[2pt] \vec{\alpha}_j \end{pmatrix}\, =\ 
\underbrace{\begin{pmatrix} (\t_j^\dagger)^{-1} & -(\t_j^\dagger)^{-1}\r_j^{\dagger} \\[2pt] -(\t_j^T)^{-1}\r_j & (\t_j^T)^{-1} \end{pmatrix}}_{\displaystyle \M_j}
  \begin{pmatrix} \vec{\beta}_{j-1} \\[2pt] \vec{\alpha}_{j-1} \end{pmatrix} .
\end{equation}
For a real and symmetric mass matrix $\sf m$, the coefficients before and after the event satisfy
\Beq
|\vec{\alpha}_j|^2-|\vec{\beta}_j|^2=|\vec{\alpha}_{j-1}|^2-|\vec{\beta}_{j-1}|^2=N_{\rm f}\,. \label{equ:Wr}
\Eeq 
The normalization in (\ref{equ:Wr}) is consistent with the assumption that the fields are free between the interactions and has the correct limit for $\Nf \to 1$. 
The transfer matrix $\M_j$ is now a $2\Nf \times 2\Nf$ matrix, whereas $\r_j$ and $\t_j$ are $\Nf\times\Nf$ matrices satisfying  ${\r}_j^\dagger {\r}_j+{\t}_j^\dagger {\t}_j= \mathbb{1}$. In addition, ${\rm Im}[\delta^{ac}V_{,cb}]=0$ implies that $\r = \r^T$.
In an explicit example the entries of the transfer matrix would be determined from matching the mode functions across the non-adiabatic scattering event (cf.~Appendix~\ref{sec:QM}). 
The total transfer matrix $\M(\Ns)$ after $\Ns$ scatterings is the same matrix product as before, cf.~eq.~(\ref{equ:MT}). The coefficients before the first scattering are then linked in a simple way to those after $\Ns$ scatterings 
 \begin{equation}
\begin{pmatrix} \vec{\beta}_{\Ns} \\[2pt] \vec{\alpha}_{\Ns}\end{pmatrix}\, =\ 
\underbrace{\begin{pmatrix} (\t^\dagger)^{-1} & -(\t^\dagger)^{-1}\r^{\dagger} \\[2pt] -(\t^T)^{-1}\r & (\t^T)^{-1} \end{pmatrix}}_{\displaystyle \M(\Ns)}
  \begin{pmatrix} \vec{\beta}_{0} \\[2pt] \vec{\alpha}_{0} \end{pmatrix} .
\end{equation}

  Away from the non-adiabatic events, we can unambiguously define an occupation number of the fields. In particular, the total occupation number after $\Ns$ events is
\Beq
n(\Ns)
&\equiv\frac{1}{2k}\left(|\delta\dot{\vec{\phi}}_{\Ns}|^2+k^2|{\delta\vec{\phi}}_{\Ns}|^2\right)-\frac{\Nf}{2}\,
=|\vec{\beta}_{\Ns}|^2\,
={\rm Tr}\left[{\vec{\beta}_{\Ns}^{\phantom\dagger}\vec{\beta}^{\hskip 2pt \dagger}_{\Ns}}\right] .
\Eeq
Assuming that we start in the vacuum state [i.e.~$\vec{\beta}_0=\vec{0}$ and  $\vec{\alpha}_0=(e^{i\delta_1},\hdots,e^{i\delta_{\Nf}})$], we get ${\rm Tr}[{\vec{\beta}_{\Ns}^{\phantom\dagger}\vec{\beta}^{\hskip 2pt \dagger}_{\Ns}}] 
={\rm Tr}[(\t^{\dagger}\t)^{-1}\r^\dagger\vec{\alpha}_0^{\phantom \dagger}\vec{\alpha}_0^\dagger\r]
$, where we have used the cyclic property under the trace operation. Note that
$\vec{\alpha}_0^{\phantom \dagger} \vec{\alpha}_0^\dagger=\I+\mathbb{O}$,
where $\mathbb{O}$ has vanishing entries on the diagonal. 
Using $\r^\dagger\r+\t^\dagger\t=\I$, we then find
\Beq
n
&={\rm Tr}\left[(\t^{\dagger}\t)^{-1}(\I+\r^\dagger\mathbb{O}\r)-\I\right]
={\rm Tr}\left[(\t^{\dagger}\t)^{-1}-\I\right] .
\Eeq
In the last step, we have again used the cyclic property of the trace and ${\rm Tr}[(\r\t^{-1})^\dagger\mathbb{O}(\r\t^{-1})]={\rm Tr}[\mathbb{O}]=0$. This motivates the following definition for the {\it {occupation number matrix}}\hskip 2pt\footnote{We have found this definition convenient both for numerical computations and for the derivation of the multi-field Fokker-Planck equation (see Appendix~\ref{sec:FP}).}

\beq
{\n} \equiv ({\t}^{\dagger}{\t} )^{-1}- \mathbb{1}\, .
\label{eq:nmatrixdef}
\eeq
The eigenvalues $n_a$ of the matrix $\n$ describe the number of particles of each field (possibly in a rotated field basis). The total number of particles is the sum over all eigenvalues
\beq
n  =  \sum_{a=1}^{\Nf}n_a={\rm Tr}[\n]\,.
\eeq
 In terms of the transfer matrix, this can be written as
 \Beq
 n=\frac{1}{2}{\rm Tr}\left[\frac{1}{4}\left\{2 \cdot \mathbb{1}+ {\M}{\M}^{\dagger}+ ({\M}{\M}^{\dagger})^{-1}\right\}-\mathbb{1}\right] . \label{equ:nM}
 \Eeq
We are interested in the statistics of the stochastic particle production.
As before, the most efficient way to derive this is as a solution to a Fokker-Planck equation.

\subsection{Fokker-Planck Equation}
\label{ssec:MultiMoments}
The multi-field generalization of the FP equation (\ref{equ:FP}) is a bit more complex, and is derived in detail in Appendix~\ref{sec:FP}. 
After some work, one finds that the evolution equation for the joint probability distribution of the eigenvalues $n_a$ is
\Beq
\label{equ:MFP}
\frac{1}{\mu_k}\frac{\partial}{\partial \tau}P(n_a;\tau)
&\ =\ \sum_{a=1}^{N_{\rm f}}\left[(1+2n_a)+\frac{1}{\Nf+1}\sum_{b\ne a}\frac{n_a+n_b+2n_an_b}{n_a-n_b}\right]\frac{\partial P}{\partial n_a}\\
&\qquad+\frac{2}{\Nf+1}\sum_{a=1}^{N_{\rm f}}n_a(1+n_a)\frac{\partial^2 P}{\partial n_a^2} \ ,
\Eeq
where we have defined the mean particle production rate as 
\beq
\mu_k \equiv \frac{1}{N_{\rm f}}\frac{\langle n \rangle_{\delta\tau}}{\delta\tau}\, .
\label{equ:muk}
\eeq
Although our derivation in Appendix~\ref{sec:FP} uses the {\it maximum entropy ansatz} (see Appendix~\ref{sec:MEA}), the form of eq.~(\ref{equ:MFP}) also follows under less restrictive assumptions~\cite{1991PhRvL..67..342M}. Without the maximum entropy ansatz (or related simplifications), the FP equation in the multi-field scenario can be significantly more complex  (e.g.~\cite{2012PhRvB..86a4205X,PhysRevB.66.115318,Douglas2014}). 

\subsection{Moments of the Density}

A formal solution to the FP equation~\eqref{equ:MFP} is provided in \cite{PhysRevLett.74.2776}. However, as before, it is convenient to transform the FP equation into a hierarchy of equations for the moments of the occupation numbers:
\begin{align}
\frac{N_{\rm f}+1}{2} \frac{1}{\mu_k} \frac{\partial \langle F\rangle}{\partial \tau} &\,=\, \left\langle \sum_{a=1}^{N_{\rm f}} \left[ n_a(1+n_a) \frac{\partial^2 F}{\partial n_a^2} + (1+ 2 n_a) \frac{\partial F}{\partial n_a}\right]\right. \nonumber \\
&\left. \ \ \ \ +\, \frac{1}{2}\sum_{a \ne b}^{N_{\rm f}} \frac{1}{n_a - n_b} \left[n_a(1+n_a) \frac{\partial F}{\partial n_a} - n_b(1+ n_b) \frac{\partial F}{\partial n_b} \right] \right \rangle\, ,  \label{equ:F}
\end{align}
where $\langle F\rangle \equiv \int \prod_a \d n_a F(n_a) P(n_a;\tau)$. A closed set of  equations for the moments of $n = \sum_a n_a$ then is
\begin{align}
\frac{1}{\mu_k}\frac{\partial \langle n \rangle}{\partial \tau} &\,=\, N_{\rm f} + 2\langle n \rangle\, , \label{equ:n1}\\
\frac{(N_{\rm f}+1)}{2} \frac{1}{\mu_k}\frac{\partial \langle n^2\rangle }{\partial \tau}  &\,=\,  (N_{\rm f}^2+N_{\rm f}+2) \langle n \rangle + 2 (N_{\rm f}+1) \langle n^2\rangle + 2 \langle n_2 \rangle\, , \label{equ:n2} \\
\frac{(N_{\rm f}+1)}{2} \frac{1}{\mu_k} \frac{\partial  \langle n_2 \rangle}{\partial \tau} &\,=\,  (2N_{\rm f} +2) \langle n \rangle +  \langle n^2 \rangle + (2N_{\rm f} +3) \langle n_2 \rangle\, ,  \label{equ:n_2}
\end{align}
where $n_2 \equiv \sum_{a} n_a^2$ (not to be confused with the $n_2$ used in \textsection\ref{subsec:FPEq}, or with $n_{a=2}$ in this section). Remarkably, these equations can be solved exactly.  
The evolution of the mean is simply $\Nf$ copies of (\ref{equ:mean}),
\beq
\langle n \rangle \,=\, \frac{N_{\rm f}}{2}\left(e^{2 \mu_k \tau} - 1 \right)  , \label{equ:nX}
\eeq
and the solution for the second moment is
\begin{align}
\langle n^2\rangle&\,=\, \frac{N_{\rm f}(2N_{\rm f}^2+N_{\rm f}+1)}{4(2N_{\rm f}+1)} -\frac{N_{\rm f}^2}{2}\, e^{2\mu_k \tau} + \frac{N_{\rm f}(N_{\rm f}+1)}{12}\, e^{4\left(\frac{N_{\rm f}+2}{N_{\rm f}+1}\right)\mu_k \tau} \label{equ:Nsquared}
\\[2pt]
&\hspace{0.5cm}  +\frac{N_{\rm f}(N_{\rm f}^2-1)}{3(2N_{\rm f}+1)}\,e^{4\left(\frac{N_{\rm f}+1/2}{N_{\rm f}+1}\right) \mu_k \tau} \, , \nonumber 
\end{align}
which reduces to (\ref{equ:var}) for $\Nf=1$.

\begin{figure*}[t] 
   \centering
    \includegraphics[scale=0.725]{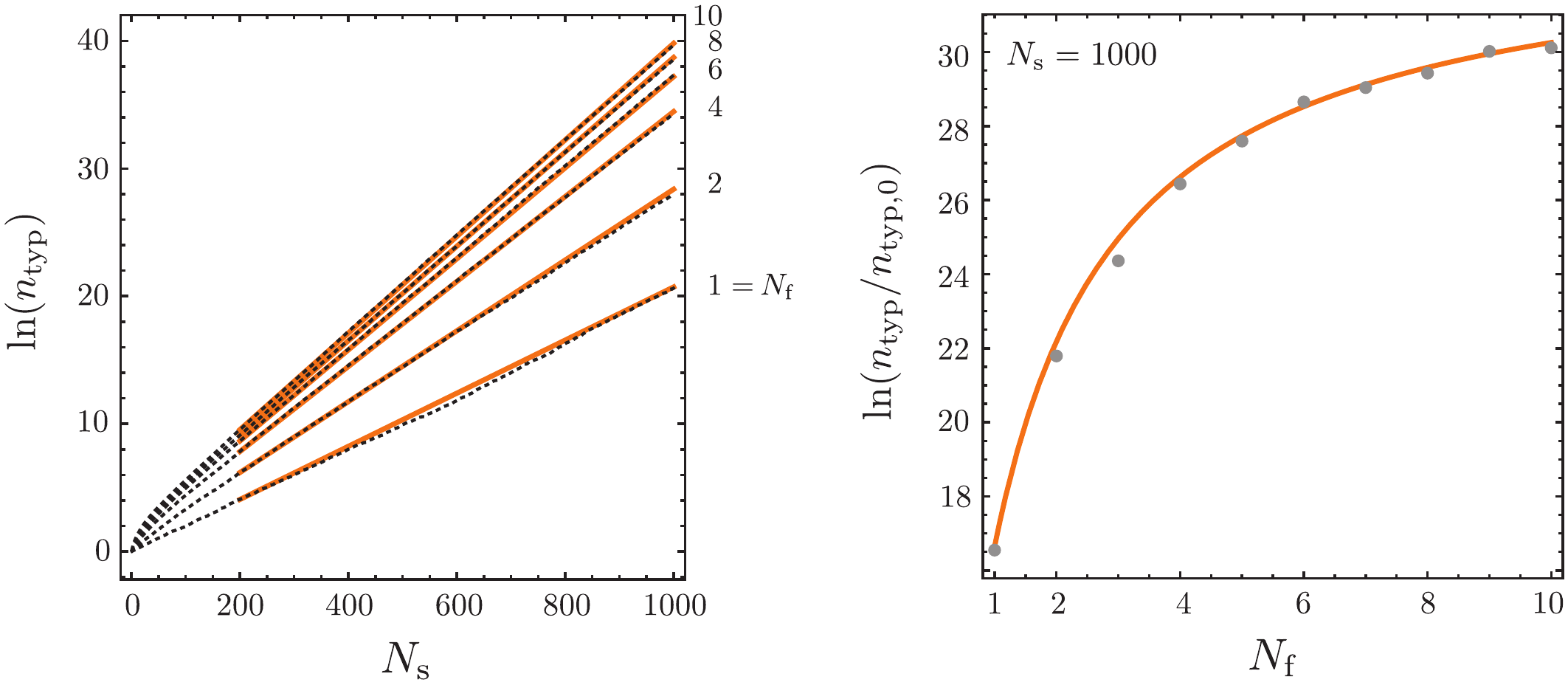} 
   \caption{Comparison between the analytic result (\ref{equ:ANALYTIC}) and our numerical simulations. 
   The ``mass matrix" for the coupled mode functions of the fields was modelled as $\m \equiv \sum_{j=1}^{N_{\rm s}}\m_j\,\delta_D(\tau-\tau_j)$, where each entry of the $N_{\rm f}\times N_{\rm f}$ matrices $\m_j$ at the random locations $\tau_j$ are drawn from uniform distributions. The strengths and distribution of the scatterers were chosen such that the mean particle production in the interval $\Delta\tau$ is $\mu_k\Delta\tau = 0.02$ (regardless of the number of fields).
    {\it Left}: Evolution of the typical occupation number $n_{\rm typ}$ with the number of scatterings $N_{\rm s}$. Each curve represents a different number of coupled fields $N_{\rm f}$. The orange lines are predictions based on our multi-field Fokker-Planck equation, whereas the dotted lines are derived from our numerical simulations. The analytical predictions are normalized at $\Ns(\tau_0)=200$. {\it Right}: The typical occupation number as a function of the number of fields, for a fixed number of scatterings $\Ns =1000$ and with $n_{\rm typ,0} \equiv n_{\rm typ}(\tau_0)$.  We see excellent agreement between our analytical prediction and the numerical results.}
   \label{fig:MultiDeltaMean}
\end{figure*}

\vskip 4pt
As in the single-field case, the variance of $n$ grows faster than its mean.  The mean is therefore not a good measure of the typical evolution.
As before, we obtain the evolution of the typical occupation number by considering the expectation value of $\ln(1+n)$. 
The first two moments satisfy
\begin{align}
 \frac{1}{\mu_k} \frac{\partial \langle \ln(1+n)\rangle}{\partial \tau} &\,=\, \left\langle  \frac{\Nf+ 2n}{1+n} - \frac{2}{\Nf+1}\frac{n+n_2}{(1+n)^2}  \right\rangle , \\[2pt]
 \frac{1}{\mu_k} \frac{\partial \langle [\ln(1+n)]^2\rangle}{\partial \tau} &\,=\, \left\langle  2 \ln(1+n) \frac{\Nf+ 2n}{1+n} - \frac{4}{\Nf+1}(\ln(1+n)-1)\frac{n+n_2}{(1+n)^2}  \right\rangle .\label{equ:log}
\end{align}
It is easy to check that these equations reduce to (\ref{equ:logn1S}) and (\ref{equ:logn2S}) in the limit $\Nf=1$.
However, this time it is not easy to find a closed set of equations. To solve  eq.~(\ref{equ:log}) exactly requires the evolution of $n_2$, which depends on the evolution of $n_3 \equiv \sum_a n_a^3$, which depends on the evolution of $n_4 \equiv \sum_a n_a^4$, etc. We can nevertheless make progress by taking the limit of late times in~(\ref{equ:log}):
\beq
\frac{1}{\mu_k} \frac{\partial \langle \ln(1+n) \rangle}{\partial \tau}  \ \xrightarrow{\ n \gg \Nf \ }\ \frac{2\Nf}{1+\Nf}\left[1+\frac{\epsilon}{\Nf}\right]  \ \xrightarrow{\ \epsilon \to 0 \ }\ \frac{2\Nf}{\Nf+1}\, ,
\eeq
\begin{wrapfigure}{r}{.4\textwidth}
\vspace{-0.75cm}
\begin{center}
\includegraphics[width=0.38\textwidth]{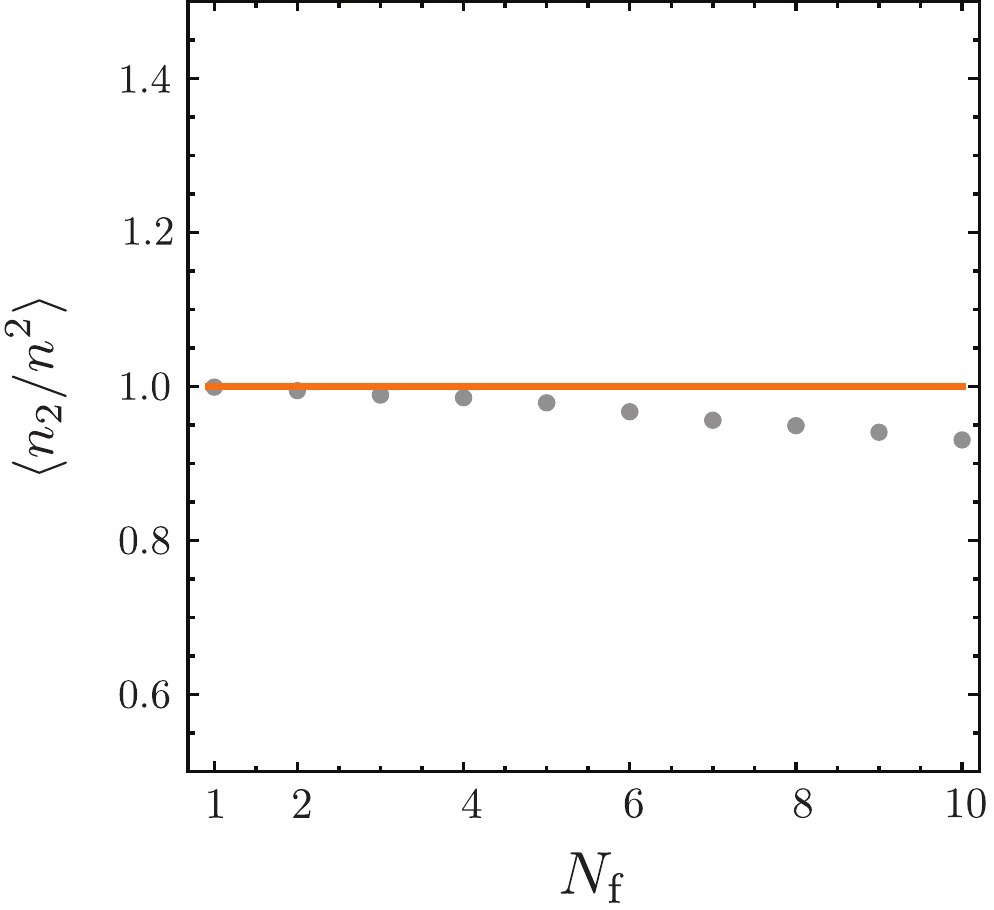}
\vspace{-0.1cm}
  \caption{Test of $\langle n_2/n^2\rangle \approx 1$ as a function of $\Nf$. The parameters of the numerical example are the same as in Fig.~\ref{fig:MultiDeltaMean}.}
  \label{fig:n2}
\end{center}
\vspace{-0.5cm}
\end{wrapfigure}
where we have defined $\epsilon \equiv 1-\langle n_2/n^2\rangle$.
In the final passage we have assumed that a single eigenvalue dominates the evolution at late times. In that case, the sum of the squares of the eigenvalues $n_a$ equals the square of the sum, and hence $n_2 \approx n^2$ (or $\epsilon \to 0$). In Fig.~\ref{fig:n2}, we demonstrate the accuracy of this approximation in a specific example. Note that the deviation from $n_2\approx n^2$ is to be compared with $\Nf$, so even a 10\% error has a small effect on the final answer. 
We then find
\begin{align}
\Delta \langle \ln (1+n) \rangle 
&= \frac{2\Nf}{\Nf+1} \mu_k(\tau-\tau_0)\, . \label{equ:lnn}
\end{align}
In Fig.~\ref{fig:MultiDeltaMean}, we show a comparison between this analytical result and a numerical example. We see that our prediction captures the functional dependence on the number of fields extremely well. 

Equation (\ref{equ:lnn}) implies that the evolution of the typical occupation number is
\beq
\boxed{n_{\rm typ}(\tau) \approx n_{{\rm typ},0}\,\exp\left[\frac{2\Nf}{\Nf+1}\mu_k(\tau-\tau_0) \right]} \ . \label{equ:ANALYTIC}
\eeq
We see that the rate of growth has a weak dependence on $\Nf$, which disappears in the limit~$\Nf \gg 1$. We also note that the late time growth of $n_{\rm typ}$ for $\Nf \gg 1$ is the square of that for $\Nf=1$. 
  
\vskip 4pt
Taking the same limits in the evolution equation~(\ref{equ:log}), we find 
\begin{align}
\frac{1}{\mu_k} \frac{\partial \langle [\ln(1+n)]^2 \rangle}{\partial \tau}  &\ \xrightarrow{\ n \gg \Nf \ }\ \frac{4\Nf}{\Nf+1} \left\langle \ln(1+n) \left[1+\frac{\epsilon}{\Nf}\right]\right\rangle  + \frac{4(1-\epsilon)}{\Nf+1}  \nonumber \\[4pt]
 &\ \xrightarrow{\ \ \epsilon \to 0 \ \ }\ \frac{4\Nf}{\Nf+1} \langle \ln(1+n) \rangle + \frac{4}{\Nf + 1} \, .
\end{align}
Substituting (\ref{equ:lnn}) and integrating, we find
\beq
\Delta {\rm Var}[\ln(1+n)]  = \frac{4}{\Nf+1} \mu_k(\tau-\tau_0)\, , \label{equ:MultiVar}
\eeq
which reduces to (\ref{equ:VarS}) in the special case $\Nf=1$.
Notice that while $\langle [\ln(1+n)]^2 \rangle$ contains a term proportional to $(\tau-\tau_0)^2$, this has cancelled in the variance. 

\subsection{Examples of Universality}
The results of this section depended only on two parameters: the mean particle production rate~$\mu_k$ and the number of fields $\Nf$. In certain limits and for certain quantities the dependence on these parameters simplifies. This corresponds to an enhanced universality of the results.
Let us give a few examples for this phenomenon:
\begin{itemize}
\item Using eqs.~(\ref{equ:nX}) and (\ref{equ:Nsquared}), we can compute the variance of the occupation number.
In the limit of a large number of fields, $\Nf \gg 1$, this becomes
\beq
{\rm Var}[n] \,=\, \frac{1}{8}\Big[1-e^{4\mu_k \tau}\left(1-4\mu_k \tau -8(\mu_k \tau)^2 \right)\Big] + {\cal O}(N_{\rm f}^{-1})\, .
\eeq
We notice that the leading term is {\it independent} of the number of fields. (The ${\cal O}(\Nf^2)$ terms in (\ref{equ:Nsquared}) have exactly cancelled against the $\Nf$-dependence of~(\ref{equ:nX}).) This universality is similar to the famous effect of universal conductance fluctuations~\cite{altshuler1985fluctuations, lee1985universal} in multi-channel wires.\footnote{In fact, the more exact analogy to universal conductance fluctuations is ${\rm Var}[n^{-1}]  =const.$ (independent of both $\Ns$ and $\Nf$).} 

\item The variance and the mean of $\ln(1+n)$ are both proportional to $\mu_k(\tau-\tau_0)$, cf.~eqs.~(\ref{equ:MultiVar}) and~(\ref{equ:lnn}). Their ratio then is  time independent and determined purely by the number of fields
\beq
 \frac{\Delta {\rm Var}[\ln(1+n)] } {\Delta \langle \ln (1+n) \rangle }= \frac{\Nf}{2} \, .
\eeq
Of course, it is not clear whether the universality of this specific ratio is physically relevant or just a coincidence.
\end{itemize}

\subsection{Random Matrix Theory}
\label{sec:RMT}

We conclude this section with a few comments on possibility of using random matrix theory~(RMT) techniques to gain further insights into the statistics of stochastic particle production. As we have seen, stochastic particle production with multiple fields can involve two large numbers: the number of fields $\Nf$ and the number of non-adiabatic events (or scatterings) $\Ns$.
Both of these large $N$'s can potentially lead to powerful applications of RMT:
\begin{itemize}
\item $\Nf$:\\  If the number of fields is large, $\Nf \gg 1$, then the random transfer matrices $\M_j$ have high dimensionality.  A lot is known about the spectrum of eigenvalues of large random matrices (see~\cite{mehta2004random} for a review).  Moreover, from (\ref{equ:nM}), we see that the eigenvalues of $\M_j \M_j^\dagger$ determine the ``local" change in the occupation number $n_j$ and then via (\ref{equ:muk}), the particle production rate $\mu_k$.  It would be interesting to use RMT to explore the probability distribution $P(\mu_k)$. 
\item $\Ns$:\\
Throughout, we have assumed that the number of scatterings is large, $\Ns \gg 1$.
In that case, the total transfer matrix $\M$ is a product of many random transfer matrices, i.e.~$\M = \prod_{j=1}^{\Ns} \M_j$. A lot is known about the asymptotic behavior of products of random matrices (see~\cite{crisanti1993products} for a review).   
Here, we highlight Oseledec's ``multiplicative ergodic theorem"~\cite{oseledec1968multiplicative}:
\begin{quote}
For $\Ns \to \infty$, the $2\Nf$ random eigenvalues $e^{\pm \nu_a}$ of ${\M} {\M}^\dagger$ tend to the non-random values $e^{\pm \gamma_a \Ns}$, with $\gamma_a$ independent of $\Ns$.
For finite $\Ns$, the $\nu_a$'s have small Gaussian fluctuations around their asymptotic limit $\gamma_a \Ns$.
\end{quote}
The parameters $\gamma_a$ in this theorem are the Lyapunov exponents. The fact that they are independent of $\Ns$ means that asymptotically the growth is purely exponential.\footnote{It is possible that exponential particle production (localization) can be avoided when the matrices $\M_j$ are drawn from ensembles belonging to certain symmetry classes \cite{2005cond.mat.11622B}.}  This is consistent with the asymptotical scalings we derived from the Fokker-Planck equation, but this time it does not assume that the scattering is weak. 
\end{itemize}
We note that some aspects of RMT (specifically for $\Ns\gg1$, but $\Nf=1$) have been used in~\cite{Zanchin:1997gf,Zanchin:1998fj} to understand the effects of noise on parametric resonance. We leave a more detailed exploration of the RMT treatment of stochastic particle production to future work. 

\section{Conclusions and Outlook}
\label{sec:Conclusions}

 In this paper, we have developed a theoretical framework for characterizing stochastic particle production in the early universe. Our approach is particularly useful when the dynamics are complex and the interactions are poorly constrained, but only coarse-grained, statistical information  
 is of interest.   In our analysis, we exploited a precise mapping between stochastic particle production in cosmology and current conduction in wires with impurities.  This allowed us to import many powerful results from the condensed matter literature  
 to the cosmological context.
  Concretely, we derived Fokker-Planck equations to describe the evolution of the particle occupation numbers and discussed their solutions. We checked that our solutions are consistent with numerical simulations of stochastic particle production in complex scenarios. We have seen hints of universality\hskip 1pt\footnote{We should caution the reader that  the universality of our results partially relies on the assumption that the interacting fields are statistically equivalent and maximally mixed with a random but uniform distribution of non-adiabatic events. Relaxing this assumption forms an interesting avenue for future work.} 
emerging in the evolution of the particle occupation numbers in systems with a large number of non-adiabatic events\hskip 1pt\footnote{Our asymptotic results apply to the limit of large occupation numbers, $n_k \gg 1$. Associated with the exponential growth of $n_k$ are an increase in the energy density, $\rho_\chi$, and the dispersion, $\langle \chi^2 \rangle$, of the $\chi$-field(s).  In a specific model, this might eventually lead  to a backreaction on both the evolution of the homogeneous background and on the dynamics of the fluctuations.  At what point these effects will be significant is model-dependent: for instance, the smaller the self-interactions of $\chi$ the longer our assumption of linearity of the equations of motion will be a good approximation. In~\cite{ Kofman:1997yn}, it was argued that large occupation numbers can occur during preheating, $10^{8} \gtrsim n_k  \gg 10^2$, without causing a large backreaction and/or a violation of the linearity assumption. We also note that by ignoring Hubble expansion in our treatment we did not capture the dilution in the occupation numbers which competes with the particle production rate.} and/or a large number of fields. We have also sketched how such universal behavior can be understood in terms of random matrix theory.

\vskip 4pt
It remains to be seen if (and how) the emergent universality in the particle production is reflected in cosmological observables.   
To analyze this, we will study the backreaction of the density of produced particles on the cosmological dynamics. 
The stress-energy of the particles can influence both the background dynamics \cite{Berera:1995ie,Green:2009ds} and the evolution of the long-wavelength curvature perturbations~$\zeta$.  As explained in~\cite{LopezNacir:2011kk}, we expect the evolution of $\zeta$ to be sourced both by a stochastic noise term (which is uncorrelated with $\zeta$) and a linear response (which is correlated with $\zeta$). Both effects need to be taken into account to derive the late-time correlation functions for~$\zeta$, and hence the effects on cosmological observables. It will be interesting to follow up on the suggestion of \cite{Green:2014xqa} that weak disorder during inflation could lead to additional noise in the power spectrum and bispectrum of $\zeta$. Finally, once the effects on the curvature perturbations have been calculated, it will also be useful to compare our analytical predictions with those of existing works involving multi-field inflation with complex potentials (e.g.~\cite{McAllister:2012am,Dias:2015rca}).

Our statistical approach may also be of interest for studies of the early stages of nonperturbative (p)reheating.  Given the complexity of the dynamics, most previous works in the preheating literature only dealt with a few components governed by relatively simple interactions. For example, the treatments in the seminal papers on the subject~\cite{Kofman:1994rk, Shtanov:1994ce, Kofman:1997yn} were restricted to daughter fields that coupled to a single inflaton field. Even in these simplified scenarios, additional stochasticity in the effective masses of the daughter fields (beyond that due to expansion) can lead to novel effects such as enhanced particle production~\cite{Bassett:1997gb, Zanchin:1997gf,Zanchin:1998fj}. Particle production in the context of {\it coupled} multi-field dynamics (which can naturally lead to stochasticity during reheating), remains relatively unexplored. Based on our present work, it is conceivable that the preheating dynamics in such complex  scenarios can be parametrized by a few effective parameters (see also~\cite{Ozsoy:2015rna}). On the more practical side, our framework will allow us to determine when a linear treatment becomes a poor approximation to the dynamics, and full nonlinear simulations are needed.  Our formalism would be an efficient way to determine the initial conditions for these simulations.
 
 \vskip 4pt
 We end with some speculative remarks. Recent observations \cite{Adam:2015rua} have revealed a remarkably simple universe. Only two numbers ($A_{\rm s}, n_{\rm s}$) are required to describe a nearly scale-free and Gaussian spectrum of adiabatic curvature perturbations~\cite{Ade:2015xua, Ade:2015lrj}.  At the same time, fundamental theories of the early universe can be quite complex.   It is therefore natural to  wonder how the simplicity of the data emerges from the apparent complexity of the underlying theories. Emergent universal behaviour  is common in condensed matter systems  and is what allows predictive power in spite of the underlying complexity of materials \cite{1972Sci...177..393A}.  It is intriguing to ask whether the simplicity of the cosmological data is similarly emergent from complexity rather than reflecting the simplicity of the underlying theory. Our framework can be thought of as a first modest step towards exploring this possibility.

\subsubsection*{Acknowledgements}

We thank Thomas Bachlechner, David Berenstein, Ed Copeland, Mafalda Diaz, Daniel Green, Raphael Flauger, Jonathan Frazer,  Matt Foster, David Kaiser,   Andrei Linde,  Enrico Pajer, David Marsh, Liam McAllister,  Rafael Porto, Christof Wetterich and Hong-Yi Xie for helpful discussions. We are grateful to the Aspen Center for Physics (NSF Grant 1066293) for its hospitality while part of this work was being carried out.  D.B.~acknowledges support from a Starting Grant of the European Research Council (ERC STG Grant 279617). M.A.~acknowledges support from a Senior Kavli Fellowship at the University of Cambridge.

\newpage
\appendix
\section{Fokker-Planck for Multiple Fields} 
\label{sec:FP}
In this appendix, we derive the Fokker-Planck equation for stochastic particle production with multiple fields.

\subsection{Polar Form of the Transfer Matrix}
\label{subsec:Matrix}

Consider the transfer matrix
\beq
\M= \begin{pmatrix} (\t^\dagger)^{-1} & -(\t^\dagger)^{-1}\r^\dagger \\[2pt] -(\t^T)^{-1}\r & (\t^T)^{-1} \end{pmatrix} . \label{equ:Mdef}
\eeq
We wish to write this in the so-called `polar form'~\cite{1988AnPhy.181..290M}. First, note that the `singular value decomposition' of the matrix $\t$ is given by  
\Beq
\t = \u\sqrt{\hat\T}\hskip 1pt \v\,,
\Eeq
where $\u$ and $\v$ are unitary matrices and $\hat \T$ is a diagonal matrix whose entries are the eigenvalues of $\t \t^\dagger$. The columns of the matrices $\u$ and $\v^T$ are the eigenvectors of $\t \t^\dagger$ and $\t^\dagger \t$, respectively. Note that the choice of $\u$ and $\v$ is not unique. 
For future convenience, we define another unitary matrix $\z=\v \u$,
such that
\Beq
\t=\u\sqrt{\hat\T} \hskip 1pt\z \u^\dagger \, .
\Eeq
As discussed in \textsection\ref{sec:PC} and \textsection\ref{sec:Prelim}, the reflection and transmission matrices $\r$ and $\t$ satisfy $\r^\dagger \r +\t^\dagger \t =\mathbb{1}$ and $\r = \r^T$. These properties allow us to write
\Beq
\r & = - \v^T\sqrt{\mathbb{1}-\hat{\T}}\hskip 1pt \v=- \u^*\z^T\sqrt{\mathbb{1}-\hat{\T}}\hskip 2pt \z\u^\dagger\, .
\Eeq
In terms of the diagonal number density matrix $\hat{\n} \equiv \hat{\T}^{-1}-\mathbb{1}$, we get
\Beq
\t&=\u\sqrt{(\mathbb{1}+\hat{\n})^{-1}}\hskip 1pt \z \u^\dagger \,,\\
\r&=- \u^*\z^T\sqrt{\hat{\n}(\mathbb{1}+\hat{\n})^{-1}}\hskip 2pt \z\u^\dagger\, . \label{equ:A5}
\Eeq
The polar form of the transfer matrix is then obtained by substituting (\ref{equ:A5}) into (\ref{equ:Mdef}),
\Beq
\M= \begin{pmatrix} \u & {\sf 0} \\[2pt] {\sf 0} & \u^* \end{pmatrix}\begin{pmatrix} \sqrt{\I+\hat \n} & \sqrt{\hat \n} \\[2pt] \sqrt{\hat \n} & \sqrt{\I +\hat \n} \end{pmatrix}\begin{pmatrix} \v & {\sf 0} \\[2pt] {\sf 0} & \v^* \end{pmatrix}
\,=\, \begin{pmatrix} \u\sqrt{\I+\hat \n}\,\z\u^\dagger & \u\sqrt{\hat \n}\,\z^*\u^T \\[2pt] \u^*\sqrt{\hat \n}\,\z\u^\dagger & \u^*\sqrt{\I+\hat \n}\,\z^*\u^T \end{pmatrix}  .
\Eeq
In the single-field case, the above definitions reduce to
\Beq
\u &= e^{i\phi}\,, \ \ \, {\hat \n}=n\,, \ \ \, \z=e^{i\theta}, \ \ \, \t =t= \sqrt{(1+n)^{-1}}\hskip 1pt e^{i\theta}\,,  \ \ \, \r= r=- \sqrt{n(1+n)^{-1}}\hskip 1pt e^{2i(\theta-\phi)} \, ,
\Eeq
and, hence, we find
\Beq
\M=\begin{pmatrix} e^{i\theta}\sqrt{ 1+n}& \ e^{i(2\phi-\theta)}\sqrt{n} \\[2pt] e^{-i(2\phi-\theta)}\sqrt{n} & \ e^{-i\theta}\sqrt{1+n} \end{pmatrix} .
\label{eq:PolarSF}
\Eeq
This is the form of the transfer matrix that we used in \textsection\ref{subsec:FPEq}.

\subsection{Derivation of the Fokker-Planck Equation}
In terms of the parameterization of the previous section, the Smoluchowski equation~\eqref{equ:SmolM} reads
\Beq
P(\{\hat \n,\z,\u\};\tau+\delta \tau)&=\int P(\{\hat \n_{1}, \z_{1},\u_{1}\};\tau)P(\{\hat \n_{2}, \z_{2},\u_{2}\};\delta \tau)\,\d\hat \n_{2}\d\u_{2}\d\z_{2}\\
&=\langle P(\{\hat \n_{1},\z_1,\u_{1}\};\tau)\rangle_{\delta \tau}\, .
\Eeq
To derive the FP equation, we proceed as in the single-field case (cf.~\textsection\ref{subsec:FPEq}), except that the algebra will be a bit more involved.
As before, we need to express the elements of the transfer matrix~$\M_1$ in terms of the elements of $\M$ and $\M_2$ (see Fig.~\ref{fig:FP}), using $\M_1=\M_2^{-1}\M $.

\vskip 4pt
First, we consider $\n_1 =(\t_1^\dagger\t_1^{\phantom\dagger})^{-1}- \mathbb{1}$ [see eq.~\eqref{eq:nmatrixdef}], which can be written as
\Beq
\n_1 =\left[{\M_1}\right]_{11}^\dagger\left[{\M_1}\right]_{11}- \mathbb{1}\, ,
\Eeq
where
\Beq
\left[{\M_1}\right]_{11}&=\u_1 \sqrt{1+\hat \n_1}\v_1=\v_2^\dagger\left[\sqrt{1+\hat \n_2}\,\tilde{\u}_2^\dagger\sqrt{1+\hat \n}-\sqrt{\hat \n_2}\,\tilde{\u}_2^T\sqrt{\hat \n}\right]\v\, ,
\Eeq
with $\tilde{\u}_2 \equiv \u^\dagger \u_2$.
The result can be expressed in the following form
\Beq
\n_1
&=\n +\v^\dagger\Delta\n\hskip 1pt\v\, ,\\
\Eeq
where
\Beq
\Delta \n
&=\sqrt{\mathbb{1}+\hat \n}\, {\tilde{\u}}_2\hat \n_2{\tilde\u}_2^\dagger\sqrt{\mathbb{1}+\hat \n}-\sqrt{\mathbb{1}+\hat \n}\,{\tilde\u}_2\sqrt{(\mathbb{1}+\hat \n_2)\hat \n_2}\,{\tilde\u}_2^T\sqrt{\hat \n}\\[2pt]
&\qquad +\sqrt{\hat \n}\,{\tilde{\u}}_2^*\hat \n_2{\tilde{\u}}_2^T\sqrt{\hat \n}-\sqrt{\hat \n}\,{\tilde{\u}}_2^*\sqrt{\hat \n_2(\mathbb{1}+\hat \n_2)}\,{\tilde{\u}}_2^\dagger\sqrt{\mathbb{1}+\hat \n}\, .
\Eeq
 Note that $\hat \n_1, \hat \n$ are diagonal, but $\n,\n_1, \Delta \n$ are not. Thinking of $\n+\v^\dagger\Delta \n\v$ as a perturbed Hamiltonian, we will use perturbation theory to find the eigenvalues of $\n_1$ in terms of the eigenvalues of $\n$ and the matrix elements of $\Delta \n$ in the basis that diagonalizes $\n$. Let the eigenvalues of the matrices $\n_1$, $\n$, $\Delta \n$ be $n_{1a},n_a$, $\delta n_a$, respectively. Perturbation theory then leads to
\beq
n_{1a} =n_a+\delta n_a\, ,
\eeq
where
\beq
\delta n_a =(\Delta \n)_{aa}+\sum_{a\ne b}\frac{(\Delta \n)_{ab}(\Delta{\n})_{ba}}{n_a-n_b}+\cdots \, .
\label{equ:n_a}
\eeq
The single-field analog of this expression does not require perturbation theory and is given exactly by \eqref{equ:delta n}.
Similar manipulations allow us to express \{$\z_1$, $\u_1$\} in terms of \{$\hat \n,\z,\u$\} and \{$\hat \n_2,\z_2,\u_2$\}. This step is cumbersome and readers interested in the gory details can find them in~\cite{mello2004quantum}. 

\vskip 4pt
 As in the single-field case, we use the {\it maximum entropy ansatz} (see Appendix~\ref{sec:MEA}) to put physical constraints on the form of  $P_2 \equiv P(\{\hat \n_{2},\u_{2}, \z_{2}\};\delta \tau)$.  This time one finds that $P_2$ is independent of $\u_2$, i.e.~$P_2=P(\{\hat \n_2,\z_2\};\delta\tau)$,\footnote{The particular form of $P_2(\{\hat \n_2,\z_2\};\delta\tau)$ is not important. The derived FP equation is more general, and does not rely on the details related to the maximum entropy ansatz. The results of our numerical simulations are also found to be insensitive to the particular choice of $P_2$.} and the {\it persistence property} yields $P=P(\{\hat \n,\z\};\tau+\delta\tau)$. Integrating over $\z$ on both sides of the Smoluchowski equation, we get
\Beq
P(\hat \n;\tau+\delta \tau)=\langle P(\hat \n_1;\tau)\rangle_{\delta \tau}\, . \label{eq:RedSm}
\Eeq
Taylor expanding the left-hand side of equation \eqref{eq:RedSm} with respect to $\delta\tau$ and the right-hand side with respect to $\delta n_a$, we find
\beq
\begin{aligned}
\partial_\tau P(n_a;\tau)=\sum_{a=1}^{N_{\rm f}}  \frac{\partial P}{\partial n_a} \frac{\langle\delta n_a\rangle_{\delta \tau}}{\delta \tau} +\frac{1}{2}\sum_{a,b=1}^{N_{\rm f}}\frac{\partial^2 P}{\partial n_a\partial n_b}\frac{\langle\delta n_a\delta n_b\rangle_{\delta \tau}}{\delta \tau}+\cdots\, .
\end{aligned}
\eeq

Our final task is to find expressions for $\langle \delta n_a\rangle_{\delta \tau}$ and $\langle \delta n_a\delta n_b\rangle_{\delta \tau}$ using equation \eqref{equ:n_a}. 
Calculating the expectation values  $\langle \delta n_a\rangle_{\delta \tau}$ and $\langle \delta n_a\delta n_b\rangle_{\delta \tau}$ is aided by properties of the unitary matrices and also by expanding in powers of $\mu\hskip 1pt \delta \tau$, where $\mu$ is the local mean particle production rate defined as
\Beq
\mu\equiv\frac{1}{\Nf}\frac{\langle {\rm Tr}[\hat\n_{2}] \rangle_{\delta\tau}}{\delta \tau}\, .
\Eeq
The result of a tedious computation is 
\Beq
 \langle \delta n_a\rangle_{\delta \tau}&=\mu\hskip 1pt \delta \tau\bigg[(1+2n_a)+\frac{1}{\Nf+1}\sum_{b \ne a}\frac{n_a+n_b+2n_an_b}{n_a-n_b}\bigg]\, ,\\
 \langle \delta n_a\delta n_b\rangle_{\delta \tau}&=\mu\hskip 1pt \delta \tau\bigg[\frac{4}{\Nf+1}n_a(1+n_a)\bigg] \delta_{ab}\, .
\Eeq
All higher-order moments vanish at linear order in $\mu\hskip 1pt \delta \tau$.
Note that $\langle \delta n_a\delta n_b\rangle_{\delta \tau} \propto \delta_{ab}$. This {\it isotropy} is a consequence of the maximum entropy ansatz.  Putting everything together, we obtain the multi-field Fokker-Planck equation
\Beq
\frac{1}{\mu}\frac{\partial}{\partial \tau}P(n_a;\tau)
&\ =\ \ \overbrace{\sum_{a=1}^{N_{\rm f}}\left[(1+2n_a)+\frac{1}{\Nf+1}\sum_{b\ne a}\frac{n_a+n_b+2n_an_b}{n_a-n_b}\right]\frac{\partial P}{\partial n_a}}^{\rm drift}\\[2pt]
&\qquad+\underbrace{\frac{2}{\Nf+1}\sum_{a=1}^{N_{\rm f}}n_a(1+n_a)\frac{\partial^2 P }{\partial n_a^2}}_{\rm diffusion}\ .
\Eeq
We made extensive use of this equation in \textsection\ref{ssec:MultiMoments}.

\newpage
\section{Maximum Entropy Ansatz}
\label{sec:MEA}
In this appendix, we briefly describe the maximum entropy ansatz for constraining the form of the probability distribution
 $P(\M_2;\delta \tau)$ used in Appendix~\ref{sec:FP}.  For a more detailed discussion we refer the reader to \cite{mello2004quantum}.
\subsection{Single-Field Case}
We first consider the single-field case.
Let us define the Shannon entropy of the probability distribution as  
\Beq
\label{equ:ShannonA}
\mathcal{S}&\,\equiv\,-\langle\ln P(\{n_2,\theta_2, \phi_2\};\delta\tau) \rangle_{\delta\tau}\\[2pt]
&\quad\, -\gamma_1\left[\langle 1\rangle_{\delta\tau}-1\right]-\gamma_2\left[\langle n_2\rangle_{\delta\tau}-\mu \hskip 1pt\delta\tau\right]+\gamma_3\left[\langle f(\theta_2)\rangle_{\delta\tau}-\Theta\hskip 1pt \delta\tau\right] ,
\Eeq
where $f(\theta_2)$ is an arbitrary function with an extremum at $e^{i\theta_2}=1$. The three constraints  in~(\ref{equ:ShannonA}) 
serve the following purposes:
\begin{itemize}
\item The first constraint simply enforces the normalization of the distribution. 
\item The second constraint fixes the local mean particle production rate
\Beq
\mu \equiv \frac{\langle n_2\rangle_{\delta\tau}}{\delta\tau}\,.
\Eeq
\item The third constraint is slightly non-trivial.  
Recall the polar form of the transfer matrix,
\Beq
\M_2&=\begin{pmatrix} e^{i\theta_2}\sqrt{ 1+n_2}& \ e^{i(2\phi_2-\theta_2)}\sqrt{n_2} \\[2pt] e^{-i(2\phi_2-\theta_2)}\sqrt{n_2} & \ e^{-i\theta_2}\sqrt{1+n_2} \end{pmatrix} .
\label{eq:PolarStrip}
\Eeq
In the limit $\delta\tau\rightarrow 0$, we expect $\M_2\rightarrow\I$, or $\{n_2\rightarrow0, e^{i\theta_2}\rightarrow 1\}$.  A  way of imposing this constraint is to assume that the expectation value of some real, positive-definite function $f(\theta_2)$ is fixed as $\delta\tau$ decreases.  In other words, the function $f(\theta_2)$ has to have an extremum at $e^{i\theta_2}=1$. An example of such a function is
\Beq
f(\theta_2)=|e^{i\theta_2}-1|^2=4\sin^2(\tfrac{1}{2}\theta_2)\,.
\Eeq
This guarantees that the probability density peaks at $\theta_2\rightarrow 0$ when $\delta\tau\rightarrow 0$. Note that $\phi_2$ is not determined by this constraint.
\end{itemize}
\noindent
Extremizing the entropy in \eqref{equ:ShannonA} yields
\Beq
P(\{n_2,\theta_2,\phi_2\};\delta\tau)=\frac{e^{-\gamma_2 n_2}}{Z(\gamma_2)}\frac{e^{-\gamma_3f(\theta_2)}}{Z(\gamma_3)} \, , \label{equ:P2}
\Eeq
where 
\begin{align}
Z(\gamma_2) &\equiv\int \d n_2\, e^{-\gamma_2 n_2} ={\gamma_2^{-1}}\, , \\
Z(\gamma_3) &\equiv\int \frac{\d\theta_2}{2\pi}\, e^{-\gamma_3f(\theta_2)} \, .
\end{align} 
Imposing that $\langle n_2\rangle_{\delta\tau}=\mu \hskip 1pt\delta\tau$ and $\langle f(\theta_2)\rangle_{\delta\tau}=\Theta \hskip 1pt\delta\tau$, we get
\begin{align}
P(\{n_2,\theta_2, \phi_2\};\delta\tau)
&=\frac{1}{\mu\hskip 1pt\delta\tau}\exp\left(-\frac{n_2}{\mu\hskip 1pt\delta\tau}\right)\frac{\exp\left[-\gamma_3(\Theta\hskip 1pt\delta\tau){f(\theta_2)}\right]}{Z(\Theta\hskip 1pt\delta\tau)}\nonumber \\[2pt]
&=P(n_2;\delta\tau)P(\theta_2;\delta\tau)\nonumber\\[2pt]
&=P(\{n_2,\theta_2\};\delta\tau)\,.
\end{align}
We see that, within the maximum entropy ansatz, the probability distribution is {\it independent of~$\phi_2$}.  This is the only fact that we need from this analysis. In particular, although we have provided an explicit form of the distribution above, we do not need these details for the derivation of the FP equation.  As an aside, we note that for the particular choice $f(\theta_2)=4\sin^2(\tfrac{1}{2}\theta)$, we have $Z(\gamma_3)=e^{-2\gamma_3}I_0(2\gamma_3)$ and
\Beq
2\left[1-\frac{I_1(2\gamma_3)}{I_0(2\gamma_3)}\right]=\Theta\hskip 1pt\delta\tau\,, \label{equ:B11}
\Eeq
where $I_n(z)$ are modified Bessel functions of the first kind. The left-hand side of (\ref{equ:B11}) is a positive-definite, monotonically decreasing function of $\gamma_3$. Hence, as $\Theta\hskip 1pt\delta\tau\rightarrow 0$, we have $\gamma_3\rightarrow\infty$ and the probability density develops a delta function at $\theta_2\rightarrow 0$. 
\subsection{Multi-Field Case}
Let us comment on the generalization of the maximum entropy ansatz to the case with multiple fields.
The Shannon entropy of the probability distribution now is  
\Beq
\label{equ:ShannonB}
\mathcal{S}&\,\equiv\, -\langle\ln P(\{\hat \n_2,\z_2, \u_2\};\delta\tau) \rangle_{\delta\tau}\\[2pt]
&\quad\, -\gamma_1\left[\langle 1\rangle_{\delta\tau}-1\right]-\gamma_2\left[\Nf^{-1}\langle {\rm Tr}[\hat \n_2] \rangle_{\delta\tau}-\mu \hskip 1pt\delta\tau\right]+\gamma_3\left[\langle f(\z_2)\rangle_{\delta\tau}-\Theta\hskip 1pt \delta\tau\right] ,
\Eeq
where $f(\z_2)$ is an arbitrary function with an extremum at $\z_2=\I$; for instance, we may have~$f(\z_2)= {\rm Tr}\left[(\z_2-\I)(\z_2-\I)^\dagger\right]$.
All constraints in (\ref{equ:ShannonB}) have the same meanings as in the single-field case. As before, the last constraint enforces that $\M_2 \to \I$ in the limit $\delta \tau\to 0$.  To see this, we consider the polar form of the transfer matrix
 \Beq
\M_2 =  \begin{pmatrix} \u_2\sqrt{\I+\hat \n_2}\,\z_2\u_2^\dagger & \u_2\sqrt{\hat \n_2}\,\z_2^*\u_2^T \\[2pt] \u_2^*\sqrt{\hat \n}\,\z_2\u_2^\dagger & \u_2^*\sqrt{\I+\hat \n_2}\,\z_2^*\u_2^T \end{pmatrix}  ,
\Eeq
which becomes the identity for $\{\hat \n_2 \to 0, \z_2 \to \I\}$. Maximizing the entropy, we find that the probability density is {\it independent of}~$\u_2$.  As in the single-field case, the persistency property holds: the convolution of two probability densities which are independent of $\u$, is itself independent of~$\u$.

\newpage
\section{Explicit Scattering Computations}
\label{sec:QM}

In this appendix, we will show how the transfer matrices and the occupation numbers are computed in concrete examples. 

\subsection{Single-Field Case}

When the wavelength of the incoming mode is much longer than the coherence interval of the non-adiabatic event, then the profile of the mass evolution, $m(\tau)$, cannot be resolved by the wave. In that limit, ``delta-function"-scatterers are a good approximation
\beq
m^2(\tau) = \sum_{j=1}^{\Ns} m_j \, \delta_D(\tau-\tau_j) \, . \label{delta}
\eeq
We write the mode function before the $j$-th scattering as
\beq
\chi_{j-1}(\tau)=\frac{1}{\sqrt{2k}}\left[\beta_{j-1} e^{i k \tau}+\alpha_{j-1} e^{- i k \tau}\right] ,
\eeq
where $\beta_{j-1}$ and $\alpha_{j-1}$ are the standard Bogoliubov coefficients. A similar expansion applies for the mode function $\chi_j$ after the scattering.  The mode functions before and after the scattering satisfy the following junction conditions 
\beq
\begin{aligned}
\chi_{j}(\tau_j)&=\chi_{j-1}(\tau_j)\, ,\\
{\chi}_{j}^{\hskip 1pt \prime}(\tau_j) &=  {\chi}_{j-1}^{\hskip 1pt \prime}(\tau_j) -m_j\chi_{j-1}(\tau_j)\, . \label{junction}
\end{aligned}
\eeq
A transfer function $\M_j$ relates the Bogoliubov coefficients before and after the scattering
\begin{equation}
\begin{pmatrix} \beta_{j} \\ \alpha_{j} \end{pmatrix}= 
 {\M}_j
  \begin{pmatrix} \beta_{j-1} \\ \alpha_{j-1} \end{pmatrix} . \label{equ:transfer2}
\end{equation}
Using (\ref{junction}), we find
\beq
{\M}_j = 
  \left[\,\mathbb{1}+i \lambda_j\begin{pmatrix}1 &e^{- 2ik\tau_j} \\ -e^{2ik\tau_j}& -1 \end{pmatrix} \right] , \quad \ \ \lambda_j \equiv \frac{m_j}{2k}\, .
\eeq
Comparison with the general transfer matrix $\M_j$ in~\eqref{equ:transfer} yields $t_j=(1-i\lambda_j)^{-1}$ and $T_j \equiv |t_j|^2=(1+\lambda_j^2)^{-1}$.  We label by $n_j$ the ``local" change in the occupation number due to the $j$-th scattering. It is related to the transmission coefficient (and hence $\lambda_j$) via
\Beq
n_j=T_j^{-1}-1 = \lambda_j^2\,.
\Eeq 
Notice that the local change in the occupation number is large only for $k \ll m_j$. Also note that a comparison with the polar form of the transfer matrix \eqref{eq:PolarSF} implies $\theta_j=\tan^{-1}(\lambda_j)$ and $\phi_j=\tan^{-1}(\lambda_j)-k\tau_j+\pi/4$.

We caution the reader that $n_j$ is {\it not} the total occupation number after $j$ scatterings. The latter we denote by $n(j)$. 
The total occupation numbers before and after the $j$-th scattering are related by
\Beq
n(j)=n(j-1) \,+\, &\lambda_j^2\left[1+2n(j-1)+2\sqrt{n(j-1)\left\{1+n(j-1)\right\}}\cos \Delta_{j}\right]\\
+\, &2\lambda_j \sqrt{n(j-1)\left\{1+n(j-1)\right\}}\sin \Delta_{j}\, ,
\Eeq
where $\Delta_j \equiv -{\rm{arg}}\left[\alpha_{j-1}\right]+{\rm{arg}}\left[\beta_{j-1}\right]-2k\tau_j$. 
Starting from vacuum initial conditions, $n(0) = 0$, we can use this formula iteratively to describe the occupation number after many scatterings.  Alternatively, we can also obtain the occupation number after $\Ns$ scatterings by first chaining together the transfer matrices $\M(\Ns)=\M_{\Ns}\hdots\M_2\M_1$ and then deriving the final occupation number via
\Beq
n(\Ns)=[\M(\Ns)]^*_{11}[\M(\Ns)]_{11}^{\phantom{*}}-1\,.
\Eeq

\begin{figure*}[t] 
   \centering
    \includegraphics[width=3.7in]{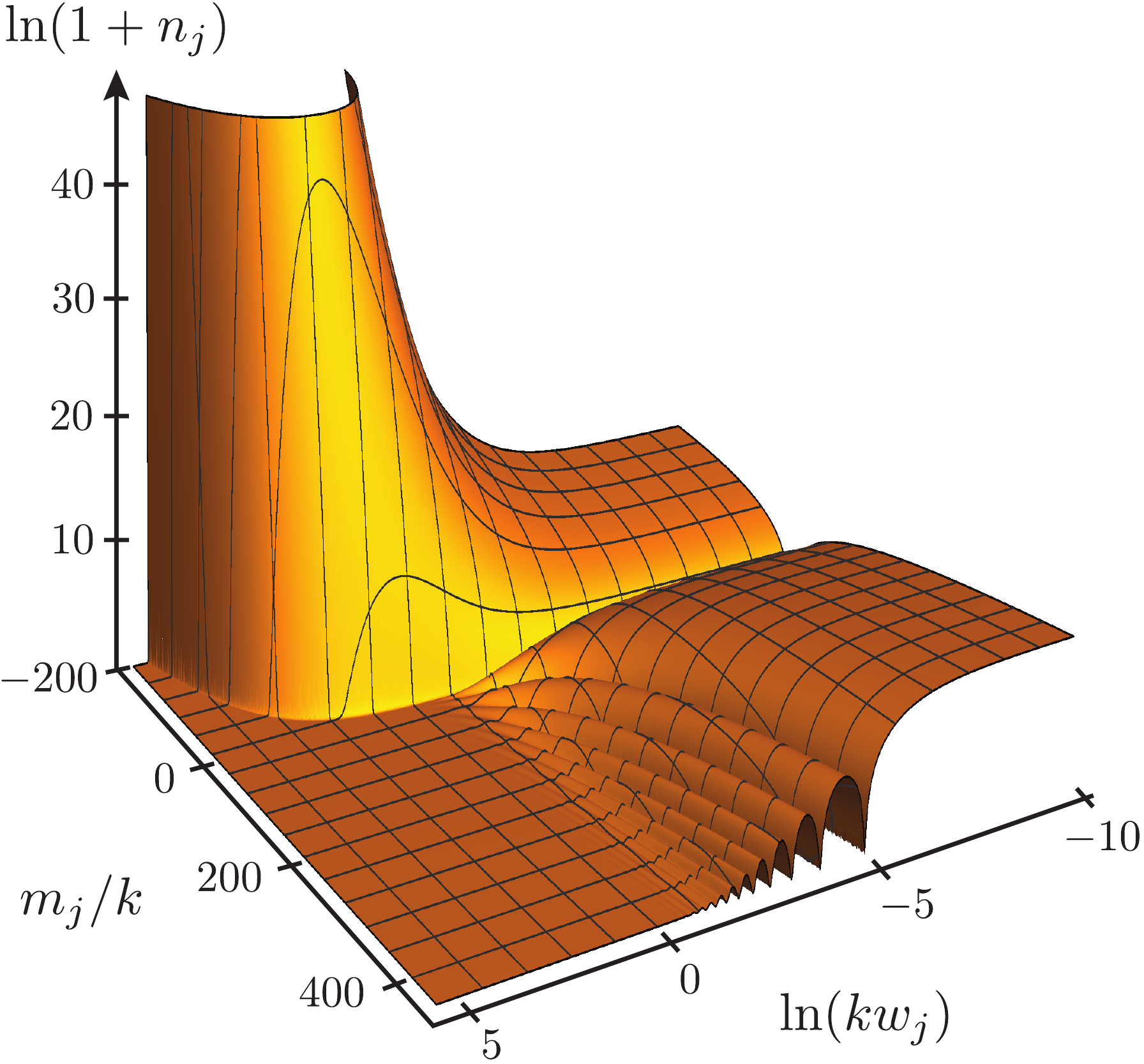}  
   \caption{Particle production for sech-scattering as a function of $k w_j$ and $m_j/k$. }
   \label{fig:TSech}
\end{figure*}
\vskip 4pt
To model a situation in which the finite ``width" (i.e.~duration) of the scattering event is relevant, we  consider ``sech"-scatterers:
\beq
m^2(\tau)=\sum_{j=1}^N \frac{m_j}{2w_j}\, \textrm{sech}^2[(\tau-\tau_j)/w_j]\, .\label{sech}
\eeq
This reduces to (\ref{delta}) in the limit $w_j \to 0$.  Using the results for transmission probabilities from~\cite{1965qume.book.....L},  
we get
\beq
n_j\equiv \frac{\cos^2(\frac{1}{2}\pi\sqrt{1+2m_j w_j}\hskip 1pt)}{\sinh^2(\pi k w_j)}\, .
\eeq
Figure~\ref{fig:TSech} shows a plot of $n_j(k,m_j,w_j)$.
We see that $n_j$ depends on
 two dimensionless ratios: $kw_j$ and $m_j/k$. The first ratio determines whether the wavelength is large or small compared to the duration of the event.  To be in the non-adiabatic regime and get significant local particle production, we require $kw_j < 1$, i.e.~the wavelength is large compared to the duration of the non-adiabatic event.  
For $m_j < 0$, there is a significant amount of particle production even with~$kw_j \sim 1$. This is the result of a temporary tachyonic instability. In the limit $w_j \to 0$ the amount of particle production is independent of the sign of $m_j$. To obtain the occupation number after $\Ns$ scatterings, we repeat the procedure described for the $\delta$-function scatterers, i.e.~we chain together the transfer matrices and read of the final occupation number from $\M(\Ns)$.

For another method to calculate particle production, see Appendix B.2 and B.3 of \cite{Chluba:2015bqa}. The authors use a {\it heat kernel} method to calculate particle production, and discuss an analytic example similar to the one discussed above.
\subsection{Multi-Field Case} 
The generalization to multiple fields is straightforward. 
For simplicity, let us restrict ourselves to $\delta$-function scatterers
and consider the following form of the coupled equations of motion for the mode functions
\beq
\ddot{\vec{\chi}}+\left[k^2\mathbb{1}+\m\right]\vec{\chi}=0\,, \quad {\rm where}\quad \m \equiv \sum_{j=1}^{N_{\rm s}}\m_j\,\delta_D(\tau-\tau_j)\,.
\eeq
The solutions between scatterings are
\Beq
\vec{\chi}_j(\tau)=\frac{1}{\sqrt{2k}}\left[\vec{\beta}_je^{ik\tau}+\vec{\alpha}_{j}e^{-ik\tau}\right] .
\Eeq
Matching these solutions at $\tau=\tau_j$, we get
\beq
\begin{pmatrix} \vec{\beta}_j  \\[2pt]  \vec{\alpha}_j\end{pmatrix}\,=\ \underbrace{\begin{pmatrix}  \left(\mathbb{1}+i\L_j\right)& ie^{-2ik\tau_j}\L_j \\[2pt] -ie^{2ik\tau_j}\L_j & \left(\mathbb{1}-i\L_j\right)\end{pmatrix}}_{\displaystyle \M_j}\, \begin{pmatrix} \vec{\beta}_{j-1}  \\[2pt]  \vec{\alpha}_{j-1}\end{pmatrix} ,
\eeq
where $\L_j\equiv \m_j/(2k)$.
Comparing this result with the general expression for the transfer matrix
\beq
\M_j= \begin{pmatrix} (\t_j^\dagger)^{-1} & -(\t_j^\dagger)^{-1}\r_j^* \\[2pt] -(\t_j^T)^{-1}\r_j & (\t_j^T)^{-1} \end{pmatrix} ,
\eeq
we find that
the transmission coefficient matrix for a single scattering event is given by $(\t_j^\dagger)^{-1}=\mathbb{1}+i\L_j$. The matrix representing the local change in the occupation numbers for that single scattering event is then given by
\Beq
\n_j = (\t_j^\dagger \t_j)^{-1}-\mathbb{1}=\L_j^2\, .
\Eeq
The trace of this matrix provides the total change in the occupation number for the scattering event. As before, we chain together the transfer matrices to obtain the result for multiple scatterings, $\M(N_{\rm s})=\M_{N_{\rm s}}\hdots \M_2\M_1$. From this we can extract $\t(N_{\rm s})$  and derive the total occupation number matrix $\n(N_{\rm s})$.

\newpage
\addcontentsline{toc}{section}{References}
\bibliographystyle{utphys}
\bibliography{Refs}

\end{document}